# Accelerated discovery of extreme lattice thermal conductivity by crystal graph attention networks and chemical bonding

Mohammed Al-Fahdi,[1] Riccardo Rurali,[2] Jianjun Hu,[3] Christopher Wolverton,[4] and Ming Hu[1,*]
[1]Department of Mechanical Engineering, University of South Carolina, Columbia, South Carolina 29208, USA
[2]Institut de Ciència de Materials de Barcelona, ICMAB–CSIC, Campus UAB, 08193 Bellaterra, Spain
[3]Department of Computer Science and Engineering, University of South Carolina, Columbia, South Carolina 29208, USA
[4]Department of Materials Science and Engineering, Northwestern University, Evanston, IL 60201, USA

## Abstract

Designing materials with targeted lattice thermal conductivity (LTC) demands electronic-level insight into chemical bonding. We introduce two bonding descriptors, namely normalized negative integrated crystal orbital Hamilton populations (-ICOHP) and normalized integrated crystal orbital bond index (ICOBI), that correlate strongly with LTC and rattling (mean-squared displacement), surpassing empirical rules and the unnormalized –ICOHP across >4,500 inorganic crystals by first-principles. We train a Crystal Attention Graph Neural Network (CATGNN) to predict these descriptors and screen ~200,000 database structures for extreme LTCs. From 367 (533) candidates with low (high) normalized -ICOHP and normalized ICOBI, first-principles validation identifies 106 dynamically stable compounds with LTC $<5$ $Wm^{-1}K^{-1}$ (68% $<2$ $Wm^{-1}K^{-1}$) and 13 stable compounds with LTC $>100$ $Wm^{-1}K^{-1}$. The descriptors' low cost and clear physical meaning provide a rapid, reliable route to high-throughput discovery and inverse design of crystalline materials with ultralow or ultrahigh LTC for applications in thermal insulation, thermoelectrics, and electronics cooling.

[*] Author to whom all correspondence should be addressed. E-Mail: hu@sc.edu

## 1. Introduction

Crystalline materials with extreme thermal conductivities, either exceptionally high or low, are technologically crucial for various applications, such as harvesting, generating, managing, and converting energy, especially thermal energy [1-5]. These technological applications mainly have two opposing aspects in terms of the lattice thermal conductivity (LTC) of involved materials. On the one hand, high LTC is desirable for heat dissipation applications such as dissipating the cumulative heat in electronic devices to extend their lifetime [6-8]. Some materials with extremely high LTC that are used for heat dissipation applications are diamond, BAs, BN, GaN, BP, and SiC [6-8]. On the other hand, extremely low LTC is desirable in thermal insulation and thermoelectric applications [4-5, 9-10]. PbTe, SnSe, $Cu_2Se$, ZrNiSn, TiNiSn, and $CoSb_3$ are some representative thermoelectrics that possess low LTC [4-5, 9-10]. However, discovering novel materials with such extreme thermal conductivities is notoriously challenging due to the fundamental competing mechanisms of the thermal transport process.

For semiconductors and insulators, phonons, i.e., the quanta of lattice vibration, are the dominant heat energy carrier and govern the LTC. According to the kinetic theory of phonon transport [1], LTC is defined as

$$k_l = \frac{1}{3} C v_g^2 \tau \quad (1)$$

where $C$ is the heat capacity, $v_g$ is the group velocity, and $\tau$ is the phonon relaxation time. Therefore, high LTC usually requires high heat capacity, group velocity, and phonon relaxation time, and the opposite holds true for low LTC. Designing new materials with desirable thermal transport properties depends on tuning those governing parameters. For example, to reduce phonon relaxation time and consequently reduce LTC in thermoelectric materials, several approaches were implemented such as introducing defects [11], rattling phonon modes [12-13], phonon softening through ferroelectric instability [14-15], and constructing materials with ions that possess lone-pair electrons [3, 16-18]. Moreover, $v_g$ is generally commensurate with $\sqrt{\frac{K}{M}}$ for isotropic materials, where $K$ can be roughly understood as bond strength, and $M$ is the atomic mass. Many phonon Boltzmann transport equation (BTE) studies have highlighted the effect of the heavy mass on reducing LTC [3-4, 19], claiming that the LTC decreases as the average mass of the materials increases, which agrees with positive correlation between the phonon group velocity and LTC. Keyes similarly expressed that high mean atomic mass in the material leads to low LTC [20]. Additionally, it was also revealed that bond stiffness plays a key role in achieving high LTC: materials with strong bonding represented by low mean squared displacement (MSD) such as carbon allotropes such as diamond tend to have high LTC [21]. Strong bonding for high LTC is also evidenced by the low p-d orbital mixing in GaAs which has higher LTC compared to the weaker bonding illustrated by high p-d orbital mixing in CuBr despite their similar mean atomic mass [22]. Slack [21, 23] had also proposed that, in order to obtain high (low) LTC, all (some of) the following criteria must be fulfilled: i) simple (complicated) crystal structure, ii) low (high) mean atomic mass, iii) low (high) anharmonicity, iv) strong (weak) interatomic bonding. Those conclusions reaffirm the criteria to tune LTC. These previous studies including both theoretical and experimental research highlight the general and qualitative physical rules for designing materials with high or low LTC for various heat transfer applications.

Despite their intuitive understanding and fast-to-deploy nature, the previously found or proposed criteria and even theoretical formula can be hardly used for accurate quantitative predictions of LTC of crystalline materials. For example, the Slack model has been widely applied for the fast evaluation of LTC with minimal time and resources, showing the great potential for high-throughput screening of LTC. However, after examining the Slack model on a large set of 353 materials, Qin *et al.* [23] found a huge discrepancy between the predicted LTC and the correspondingly measured LTC in experiments for some materials in addition to the systematical overestimation trend of LTC by the Slack model. On the other hand, the general approach of density functional theory (DFT) coupled with phonon BTE for predicting comprehensive phonon properties including LTC was developed a decade ago and has been widely applied to various materials including metals, semiconductors, and insulators [24-27] with great accuracy when compared with experiments [28-29]. This computational framework is established within the particle-like phonon transport picture by explicitly capturing the potential energy landscape near atoms' equilibrium positions, technically dubbed second-, third-, and even higher-order interatomic force constants (IFCs), which are the source of harmonic and anharmonic nature in the crystals. Despite the high accuracy of such DFT + BTE method, given the costly nature of DFT, the calculation of atomic forces for the IFCs is an extremely time-consuming process as a single material requires evaluation of several hundred, even thousands of supercells containing displaced atoms by high resolution DFT calculations. The unbearable computational cost renders the immediate deployment of such an approach impossible in a high-throughput manner to large amount of unexplored materials. This calls for urgent need of developing fast and accurate methodology for quantitatively predicting LTC of crystals and discovering novel thermal materials.

Artificial intelligence (AI) and machine learning (ML) are transforming science and engineering and poised to transform discovery and innovation. With the advent of ML algorithm usages in materials science in recent years, they have shown tremendous success in predicting various materials properties with high resolution training data fed such as by DFT calculations, including but not limited to mechanical [30-33], thermal [7, 34-39], thermoelectric [40], magnetic [41], and optical [42] properties. Despite these achievements and progress, existing ML models for LTC were trained on a limited number of structures, making them questionable when it comes to their deployment to more diverse unexplored materials, considering that the material space, such as compositions and symmetries, is huge. Moreover, lots of machine learning potentials (MLPs) were trained to replace the high cost DFT calculations and then couple with phonon BTE solutions to predict LTC values [43-44]. MLPs+BTE approach is currently restricted to a few materials or families and thus the trained MLPs cannot be easily transferred to other materials or systems. Therefore, a more sophisticated AI/ML model that can cover broad material families for quickly screening target LTCs, in particular extremely high or low LTCs, is needed.

In this work, we first propose two new chemical bonding descriptors for modeling LTC, namely negative normalized integrated crystal orbital Hamilton population (or normalized -ICOHP) and normalized integrated crystal orbital bond index (or normalized ICOBI). Their strong correlation to both LTC and mean squared displacement (MSD), another important physical property that has strong correlation with LTC, were identified based on the high accuracy DFT data of over 4,500 materials. These two descriptors are proposed for the first time for efficient screening of materials with desired LTCs, which can further enhance our understanding of harmonic and anharmonic phonon transport properties. To take advantage of these two new material descriptors for accelerated thermal material discovery, we further developed a new graph neural network (GNN) model dubbed Crystal Attention Graph Neural Networks (CATGNN) to replace DFT calculations for high-throughput prediction of our newly discovered chemical bonding descriptors. With the trained CATGNN model we predicted our chemical bonding descriptors for around 200,000 new materials to screen potentially ultralow and high LTC materials. We selected 367 (533) candidates with low (high) normalized -ICOHP and normalized ICOBI as low (high) LTC. We first predicted the stability of the materials by our separately trained from scratch CHGNet model, then some of the structures were predicted to be stable through DFT. The success rate of our CHGNet model predicting stable structure is roughly ~80%. We find that all 106 stable materials with low normalized -ICOHP and normalized ICOBI have low LTC (less than 5 W/mK), while all 15 stable materials with high normalized -ICOHP and normalized ICOBI possess high LTC. We believe that our work not only accelerates the search for materials with desirable LTC but also sheds light on electronic level descriptors to enhance our physical and chemical intuition in understanding LTC. We also believe that such work can be implemented in the inverse design of novel materials [45-46] with desirable LTC by screening materials based on our proposed intuitive chemical bonding principles.

## 2. Results and Discussion

### a) *Workflow*

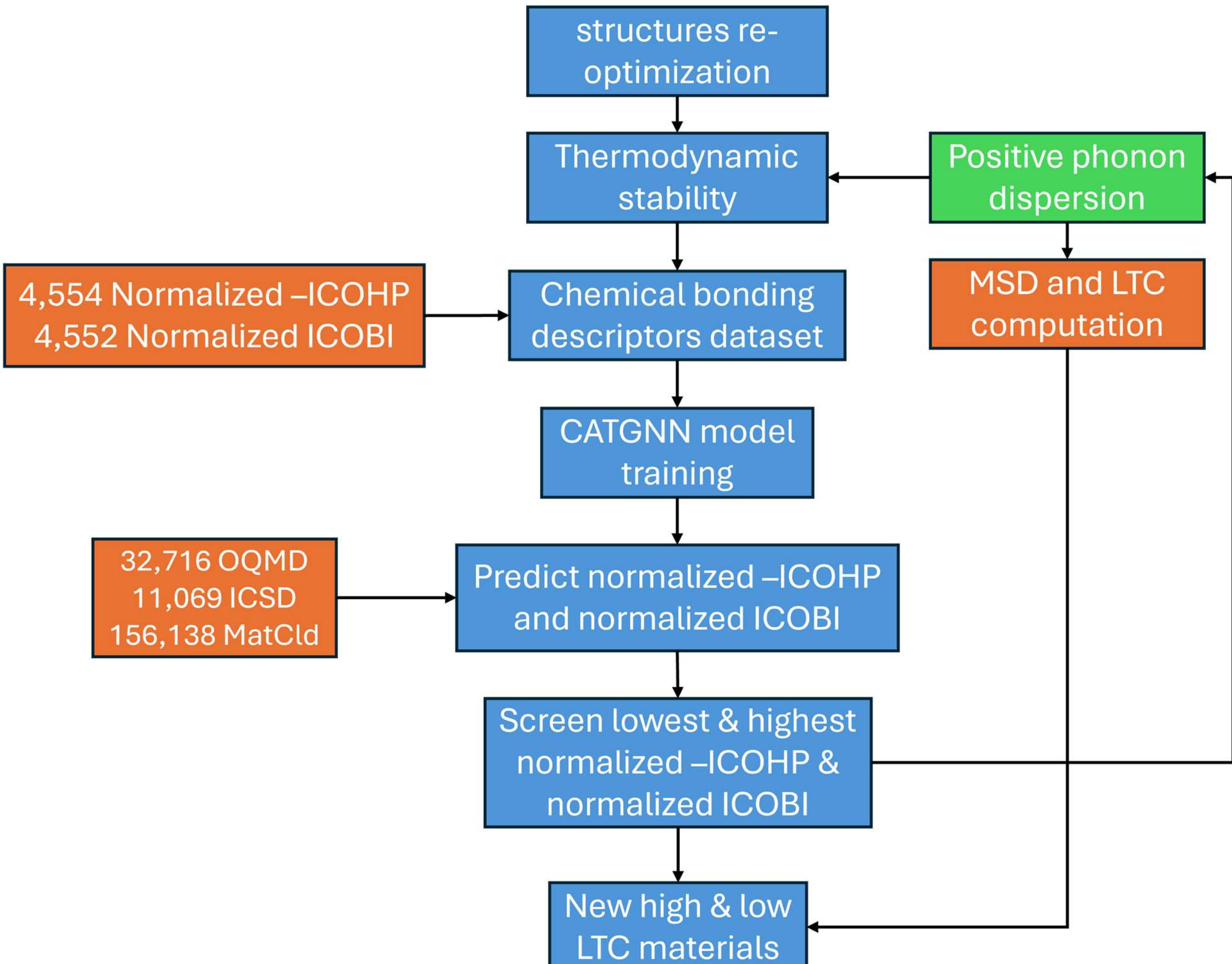

**Figure 1:** Schematic of workflow implemented in this study, which is composed of (1) structure re-optimization by our own first-principles parameters, (2) thermodynamic and dynamic stability screening, (3) chemical bonding descriptors calculations, (4) crystal attention graph neural network (CATGNN) model training, (5) deployment of trained CATGNN model to predict

~200,000 untested dataset, and (6) first-principles validation on selected 900 materials with low and high lattice thermal conductivity.

Figure 1 shows the workflow implemented in this work. Most of the structures are imported directly from the Open Quantum Materials Database (OQMD) [47], and some were taken from Refs [45-46]. We start by re-optimizing structures and computing the phonon dispersions of 4,777 materials to confirm their dynamic stability by screening the materials without imaginary frequencies in the Brillouin zone. Then, LTC of the dynamically stable structures is calculated. Next, we calculated our novel chemical bonding descriptors, namely normalized -ICOHP and normalized ICOBI, and we were able to obtain 4,554 materials with normalized -ICOHP and 4,552 materials with normalized ICOBI. The generated dataset is used for our developed CATGNN model training. The model is subsequently used to predict normalized -ICOHP and normalized ICOBI of 32,716, 11,069, and 156,138 new non-zero bandgap materials from OQMD, Inorganic Crystal Structure Database (ICSD) [48-49], and materials cloud (MatCld) [50], respectively. All structures from OQMD and ICSD databases are re-optimized separately by our own DFT parameters (see below). Then, 533 (367) materials with the lowest (highest) normalized -ICOHP and normalized ICOBI are down selected to validate their LTCs. First, we check the dynamic stability by Crystal Hamiltonian Graph Neural Network (CHGNet) model. The CHGNet model was trained from scratch by our ~116,000 separate supercell structures with random atomic displacements from equilibrium positions. These high accuracy DFT data calculated on supercells were collected during training and developing various ML models for phonon transport properties from our previous works [34-36]. Once the dynamic stability is confirmed by our new CHGNet model, i.e., no imaginary frequencies are present in the Brillouin zone, we follow by calculating their IFCs by DFT and then LTCs by solving phonon Boltzmann transport equation (BTE).

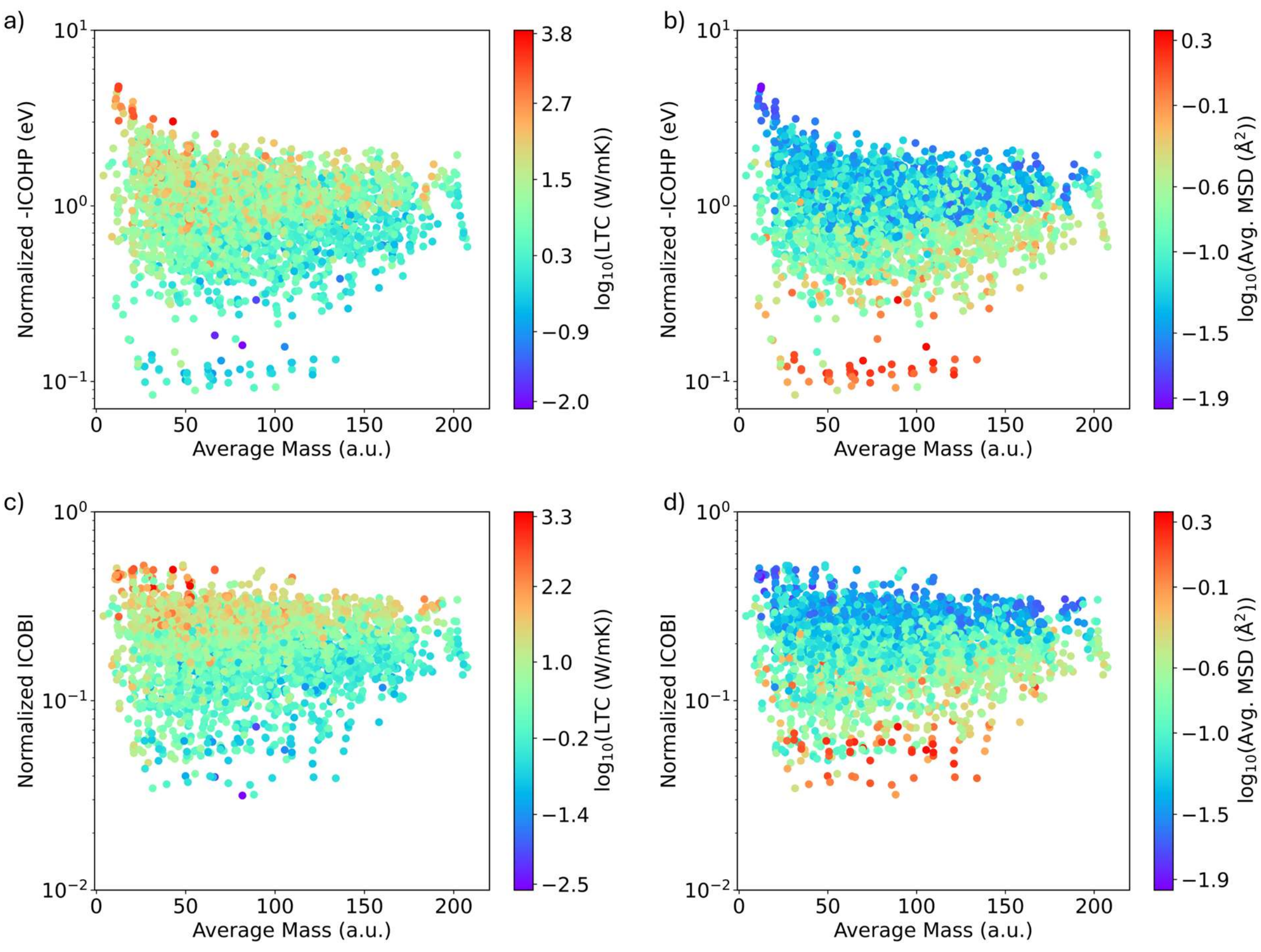

**Figure 2:** (a, b) Normalized -ICOHP vs. average mass color mapped with (a) $\log_{10}$(LTC) and (b) $\log_{10}$(Avg. MSD). (c, d) Normalized ICOBI vs. average mass color mapped with (c) $\log_{10}$(LTC) and (d) $\log_{10}$(Avg. MSD). LTC decreases (increases) as normalized -ICOHP or normalized ICOBI increases (decreases) for the same average mass. The average MSD decreases (increases) as normalized -ICOHP or normalized ICOBI increases (decreases) without significantly visible effect from the average mass to the average MSD.

*b) Normalized -ICOHP and normalized ICOBI vs. average mass correlations with LTC and MSD*

Figure 2a shows normalized -ICOHP against the average mass with $\log_{10}$(LTC) as color bars. High LTC materials exist at the top left region of the plot where high normalized -ICOHP and low average mass materials manifest. This observation confirms that materials with high LTC generally have low average mass according to Slack's model [22]. Similarly, materials with low LTC have low normalized -ICOHP. These results demonstrate that our descriptors successfully distinguish LTCs for materials with similar average mass when visualizing the plot along a vertical line (corresponding to fixed or specific average atomic mass), which cannot be realized by Slack's model, meaning that our descriptors outperform the empirical model. We also observe that even materials with the same high average mass, the LTC, even if it is low, increases with higher normalized -ICOHP and vice versa. This trend clearly explains the direct proportionality between LTC and our discovered chemical bonding strength descriptor, i.e., normalized -ICOHP. This direct correlation can be explained by the deeper potential well for interatomic bonds in materials with high normalized -ICOHP, which results in lower anharmonicity and consequently higher LTC. Moreover, it has already been demonstrated that rattling effect has inverse correlation with LTC [12-13]. In Figure 2b we present normalized -ICOHP against the average mass color mapped with $\log_{10}$(Avg. MSD) to characterize the rattling effect of materials. It is shown that the average MSD increases as normalized -ICOHP decreases and vice versa. However, average mass does not seem to have a noticeable effect on the average MSD, regardless of the value of the normalized -ICOHP. The only observed significant factor in determining the average MSD is normalized -ICOHP. The high MSD happens when normalized -ICOHP is low, which can be explained by the shallow potential well or flat potential energy landscape induced by the weak interatomic bonding in the material, which causes higher atomic displacements at given thermal energy level compared to deeper potential wells. It is interesting to observe from Figure 2b that, high average MSD and correspondingly highly anharmonic materials expand wide range of average atomic mass, meaning that the strong phonon anharmonicity could occur in many different atomic species, not just on those heavy elements as previously thought.

Furthermore, Figure 2c and 2d show the normalized ICOBI against the average mass with $\log_{10}$(LTC) and $\log_{10}$(Avg. MSD) as color bars. Materials with high LTC occur at the top left of the plot where low average mass and high normalized ICOBI materials occur. For the same or similar low average mass, LTC decreases as normalized ICOBI decreases as well, and the opposite is true when normalized ICOBI increases. Due to the higher covalent tendency in materials with higher normalized ICOBI, given that covalent bonding are strong bonds [51], the above results are fathomable. Higher average mass leads to lower LTC as stated previously by Keyes [20] and Slack model [23]. However, for the same or similar high average mass value, as normalized ICOBI increases the LTC increases as well, although still low, and vice versa. It can be observed from Figure 2d that the average MSD increases as normalized ICOBI decreases and vice versa. The average mass does not seem to have a noticeable effect on the average MSD which is same as Figure 2b. The explanation is that materials with low normalized ICOBI tend to form bonds with lower covalency and consequently weaker bonding, since covalent bonds are generally stronger than ionic and metallic bonds [51]. Therefore, the materials with higher normalized ICOBI have deeper potential wells which indicates that their atoms move around equilibrium positions with shorter displacements at given thermal energy, while atoms in materials with lower normalized ICOBI move with larger displacements. Overall, the LTC and the average MSD trend for normalized ICOBI with respect to average mass in Figure 2c and 2d is the same as the trend observed in Figure 2a and 2b.

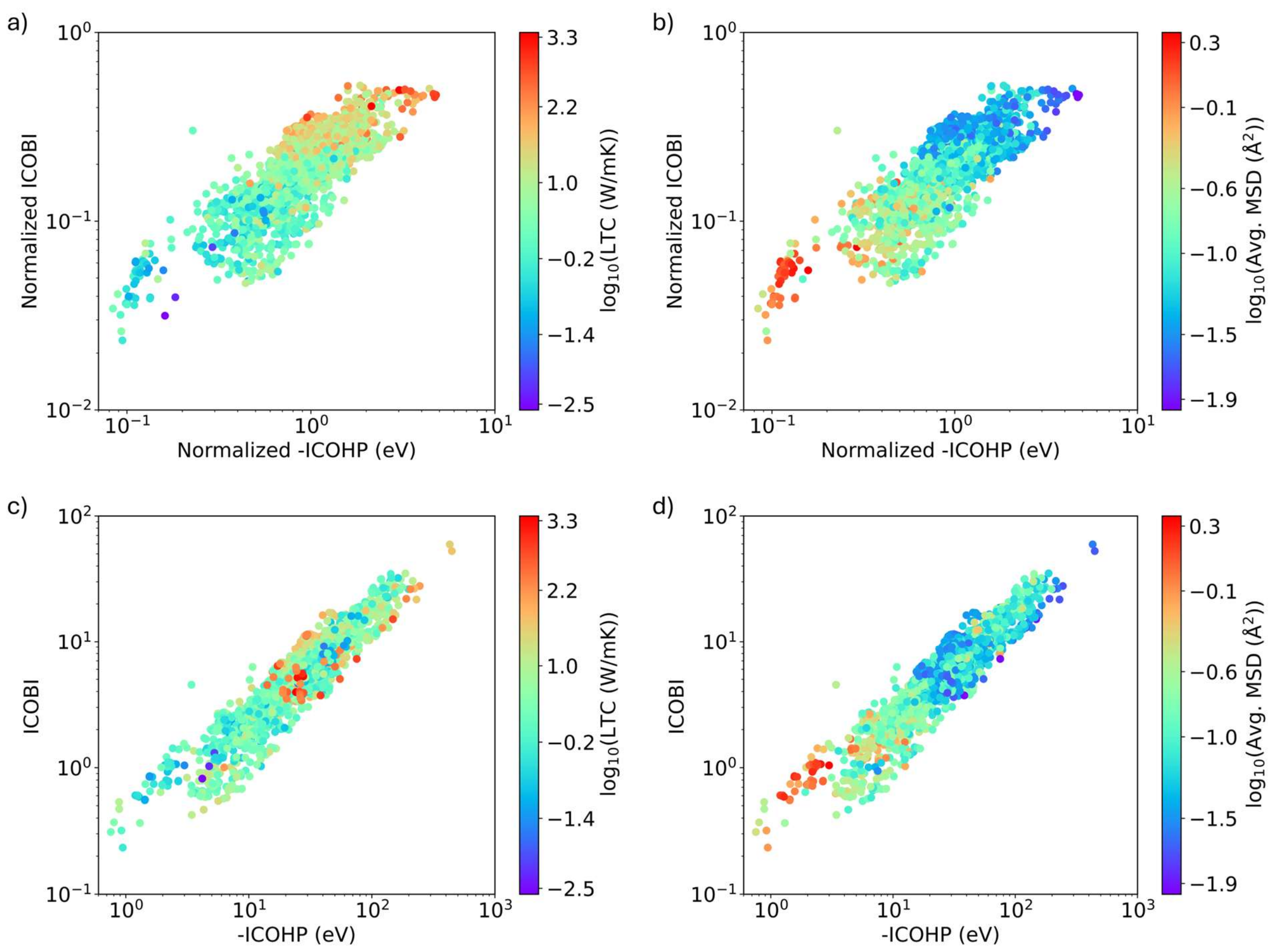

**Figure 3:** (a, b) Normalized ICOBI vs. normalized -ICOHP color mapped with (a) $\log_{10}$(LTC) and (b) $\log_{10}$(Avg. MSD). (c, d) ICOBI vs. -ICOHP color mapped with (c) $\log_{10}$(LTC) and (d) $\log_{10}$(Avg. MSD). Strong correlations between LTC, MSD and our defined descriptors are observed in top panels, while high and low LTC and MSD are scattered in the middle region of the bottom panels, indicating weak correlation between LTC, average MSD and those properties, and therefore ICOBI and -ICOHP solely are not good descriptors to screen extreme LTCs.

### *c) Correlation of normalized -ICOHP and normalized ICOBI with LTC and MSD*

Now, we explore the correlation of our newly defined normalized -ICOHP and normalized ICOBI descriptors with LTC and the average MSD directly. As shown in Figure 3a, as both normalized -ICOHP and normalized ICOBI increases, the LTC increases which illustrates the positive direct correlation between LTC and the two normalized descriptors. The results can be explained by the deep potential well in the interatomic bonds with high normalized -ICOHP and normalized ICOBI, which decreases anharmonicity and increases LTC consequently. Moreover, bonding strength is directly proportional with phonon group velocity which is directly proportional with LTC as shown in Equation (1). The characterization of LTC from these proposed chemical bonding strength descriptors is novel and has not been reported before. In Figure 3b we also report for the first time that our descriptors have inverse (negative) correlation with the average MSD, meaning that as normalized -ICOHP and normalized ICOBI increase, the average MSD decreases. The analysis and results from Figure 3 offer deep understanding and insights into LTC and MSD from chemical bonding principles. Considering the cheap computational cost of calculating our newly defined normalized -ICOHP and normalized ICOBI, these two chemical bonding descriptors offer a quick route to screen ultralow (strong anharmonicity) or high (weak anharmonicity) LTC materials by evaluating or predicting their chemical bonding characteristics, which would be much beneficial for designing novel crystalline materials with extreme LTCs for phonon-mediated applications.

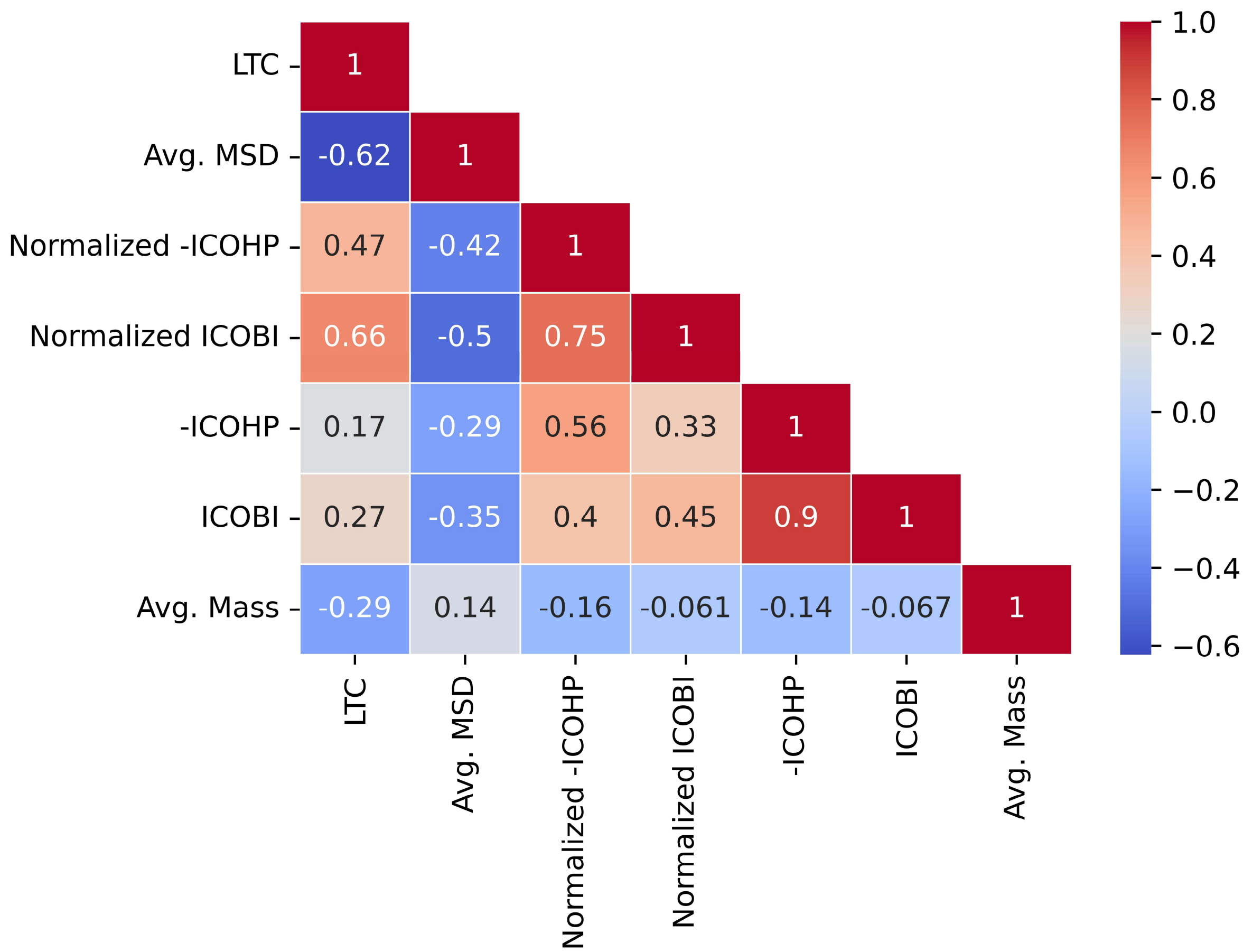

**Figure 4:** Pearson correlation among lattice thermal conductivity (LTC), average mean squared displacement (MSD), normalized -ICOHP, normalized ICOBI, -ICOHP, ICOBI, and average mass of the structure.

It is worth comparing our proposed descriptors with the sole ICOHP quantity, i.e., without normalization, which was used to explain thermal transport properties in recent studies [22, 52-55]. It was reported that antibonding states in COHP due to p-d orbital mixing leads to phonon anharmonicity [22] due to forming filled p-d bonding states and p-d* antibonding states below the Fermi energy [22, 53-55]. This phenomenon is observed in CuBr, ZnSe, and GaAs which approximately have the same mean atomic mass, but have varying LTCs of 1.25, 19, and 45 W/mK, respectively [22]. Studies on a few materials [22, 52, 55] reported that the low LTC in their systems is due to the antibonding states in COHP. Moreover, Yuan *et al.* [56] attributed the cause of ultralow LTC to the existence of antibonding valence bonds in COHP for many binary materials that contained K, Rb, or Cs. These previous studies [22, 52-56] on a limited number of materials might give the community a possible misleading impression that the existence of antibonding states of COHP could always lead to strong anharmonicity and thus the existence of COHP antibonding states might be used as a general descriptor for screening low LTC among a wide range of materials with various compositions. In this study, we found that the existence of antibonding states in COHP is not adequate to screen for low LTC materials. We utilize -ICOHP and ICOBI as chemical bonding strength descriptors and plot them with LTC and average MSD as color mapping in Figure 3c and 3d, respectively. Lots of red/orange and purple/blue dots, corresponding to high LTC and low average MSD in Figure 3c and 3d, respectively, appear in the middle region of the plot, meaning that the two bonding chemical strength descriptors do not have a unique correlation with LTC and average MSD and therefore are not good enough to serve as descriptors to screen materials with extreme LTCs and MSDs. Compared with our proposed descriptors presented in Figure 3a and 3b, the normalized -ICOHP and normalized ICOBI descriptors are better descriptors. For instance, Figure 3b illustrates the physically expected MSD trend, i.e., as bonding strength descriptors of normalized -ICOHP and normalized ICOBI increase, the average MSD decreases. This proves that such normalization treatment is necessary to represent various materials classes with varying LTC and average MSD values, as it properly considers materials with the same reduced formula but different stoichiometries in the primitive cell formula due to different phases. It is worth noting that COHP and COBI are intensive properties. However, normalization is necessary due to the accumulation of the ICOHP and ICOBI of all the bonds in the materials of which the number of bonds is not equivalent. The following example of BAs with hexagonal and cubic phases corroborates the necessity of normalization. For example, BAs has two phases: cubic (space group no. 216) and hexagonal (space group no. 186) with primitive cell formulas of BAs and $B_2As_2$, respectively. The projected orbitals of B atom are 2s and 2p, and the projected orbitals of As atom are 4s, and 4p. Since the hexagonal phase has twice more atoms and therefore twice more projected orbitals than the cubic phase, it is no surprising that the hexagonal phase of BAs has higher -ICOHP and ICOBI (-ICOHP and ICOBI are approximately twice as high in the hexagonal phase (48.5 eV and 3.961, respectively) than the cubic one (24.1 eV and 7.891, respectively)). However, upon looking at the normalized -ICOHP and normalized ICOBI, the values are approximately the same (normalized -ICOHP of 3.032 eV and 3.020 eV and normalized ICOBI of 0.493 and 0.495 for hexagonal and cubic phases, respectively). The difference in normalized -ICOHP and normalized ICOBI can be attributed to the difference in the different number of nearest neighbors, bond lengths, local environment in each site due to the change in phase which caused the difference in the bonding interactions in both phases. Other examples can be found in Table S1 in Supplemental Information. Another example is $Mg_8Cd_8S_8O_8$ with LTC of 1.8 W/mK. The -ICOHP of $Mg_8Cd_8S_8O_8$ (158.8 eV) is roughly 6.5 times higher than cubic BAs, but its normalized -ICOHP (0.946 eV) is at least 3 times lower, which manifests its ultralow LTC as compared to cubic BAs. These examples further substantiate that the normalization treatment properly considers several phases and various materials classes with varying number of atoms. We also calculated Pearson correlation coefficients of normalized -ICOHP and normalized ICOBI when correlated with LTC shown in Figure 4. Our proposed descriptors show significantly higher Pearson correlation (0.47 and 0.66 for normalized -ICOHP and normalized ICOBI, respectively) than the sole -ICOHP and ICOBI (0.17 and 0.27). As a side note, Pearson correlation of average mass with LTC is -0.28 which is lower in magnitude than our proposed descriptors, which demonstrates that our descriptors are more correlated with LTC than the average mass that has been largely used in various studies to screen low LTC materials [3-4, 19-23]. Figures 2-4, Table S1, and Figure S1 all demonstrate that our proposed descriptors are more universal for a wide range of materials than previous studies and offer chemical bonding strength insights into LTC and MSD.

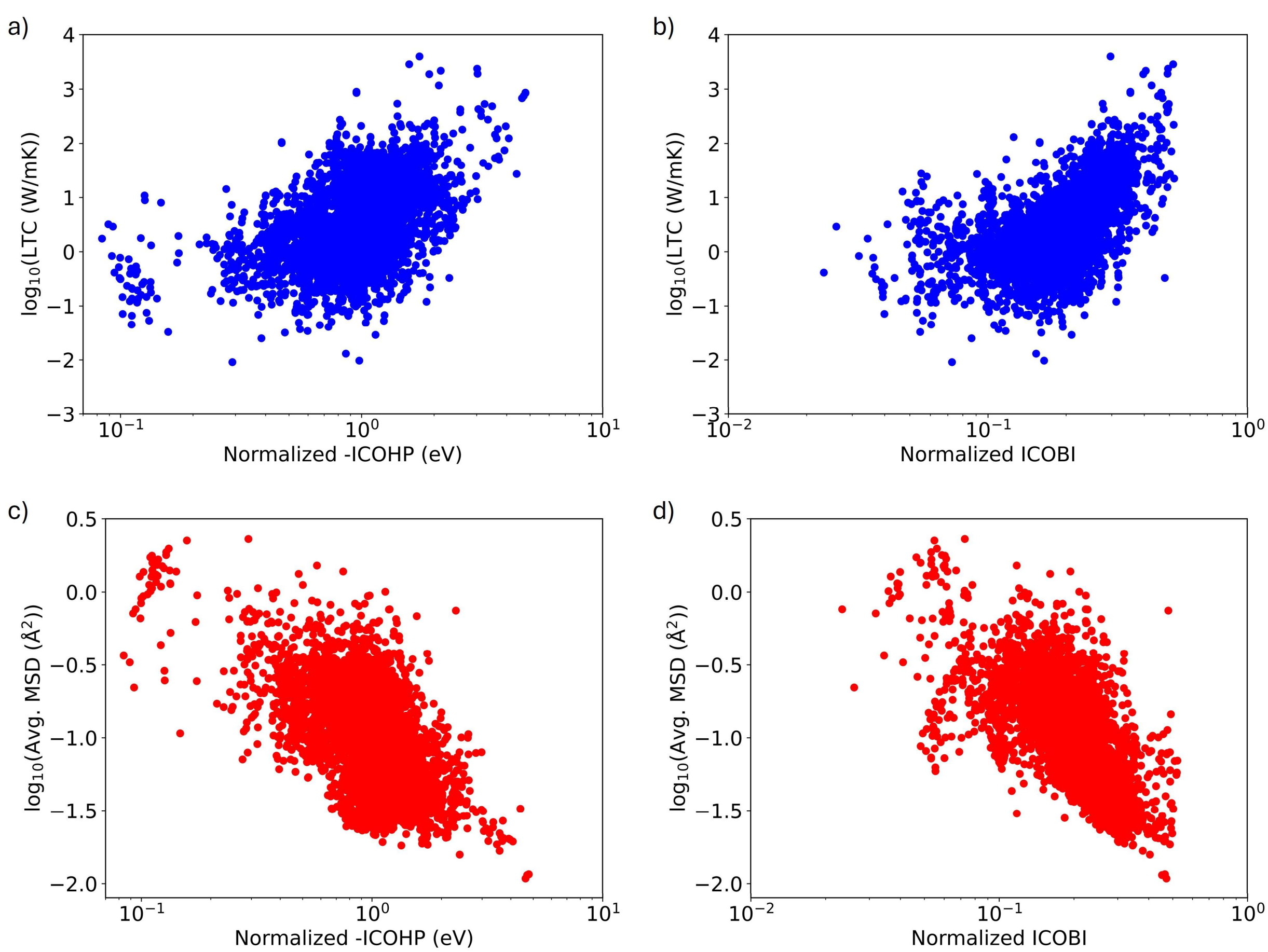

**Figure 5:** Plots of a) $\log_{10}$(LTC) vs. normalized -ICOHP, b) $\log_{10}$(LTC) vs. normalized -ICOBI, c) $\log_{10}$(MSD) vs. normalized ICOHP, and d) $\log_{10}$(MSD) vs. normalized ICOBI. These plots reaffirm the direct correlation of LTC with normalized -ICOHP and normalized ICOBI. Also, they reaffirm the inverse correlation of MSD with normalized -ICOHP and normalized ICOBI.

**Table 1:** Average values of LTC and average MSD values for arbitrary selected high and low normalized -ICOHP and normalized ICOBI.

| **Conditions** | **Average LTC ($Wm^{-1}K^{-1}$)** | **Average MSD (Å$^2$)** |
|---|---|---|
| Normalized -ICOHP > 2 eV & Normalized ICOBI > 0.3 | 240.5 | 0.0523 |
| Normalized -ICOHP > 2 eV (only) | 139.85 | 0.04907 |
| Normalized ICOBI > 0.3 (only) | 71.01 | 0.04733 |
| Normalized -ICOHP < 0.4 eV & Normalized ICOBI < 0.1 | 1.15 | 0.724 |
| Normalized -ICOHP < 0.4 eV (only) | 1.33 | 0.5775 |
| Normalized ICOBI < 0.1 (only) | 2.014 | 0.3979 |

For further observation of LTC and average MSD trends seen in Figures 2-3, we plot LTC and MSD vs. our two proposed descriptors in Figure 5a to 5d, respectively. We observe that LTC has a direct correlation with the normalized -ICOHP and normalized ICOBI, while inverse correlation occurs between MSD and normalized -ICOHP and normalized ICOBI as also seen in Figures 2-3. We also conduct another statistical approach using our chemical bonding descriptors. In this approach, we calculate the average properties of LTC and average MSD of various materials utilizing arbitrary high (low) normalized -ICOHP and normalized ICOBI as shown in Table 1. Those values are arbitrarily and visually selected from the original Figure 3 by estimating the high (low) LTC and MSD regions. The average LTC of the materials with normalized -ICOHP > 2 eV and normalized ICOBI > 0.3 is 240.5 W/mK and 0.0523 Å2, respectively. However, if the sole normalized -ICOHP > 2 eV or normalized ICOBI > 0.3 are utilized to get the average LTC of the materials, 139.85 W/mK and 71.01 W/mK are obtained, respectively. That indicates that using both chemical bonding descriptors potentially yields higher LTC compared to the single descriptor. Moreover, according to Table 1 the average properties of LTC and average MSD of the materials with normalized -ICOHP < 0.4 eV and normalized ICOBI < 0.1 is 1.15 W/mK and 0.724 Å2, respectively. However, if the sole normalized -ICOHP < 0.4 eV or normalized ICOBI < 0.1 is utilized, the average LTC of those materials becomes 1.33 W/mK and 2.014 W/mK, respectively. Those results are not significantly different from utilizing both descriptors which possibly indicates that using both descriptors has more impact on the higher LTC region but does not significantly affect the low LTC region. The above results demonstrate that those descriptors can be used from the 2D map of normalized -ICOHP and normalized ICOBI to estimate high and low LTC and average MSD materials regions.

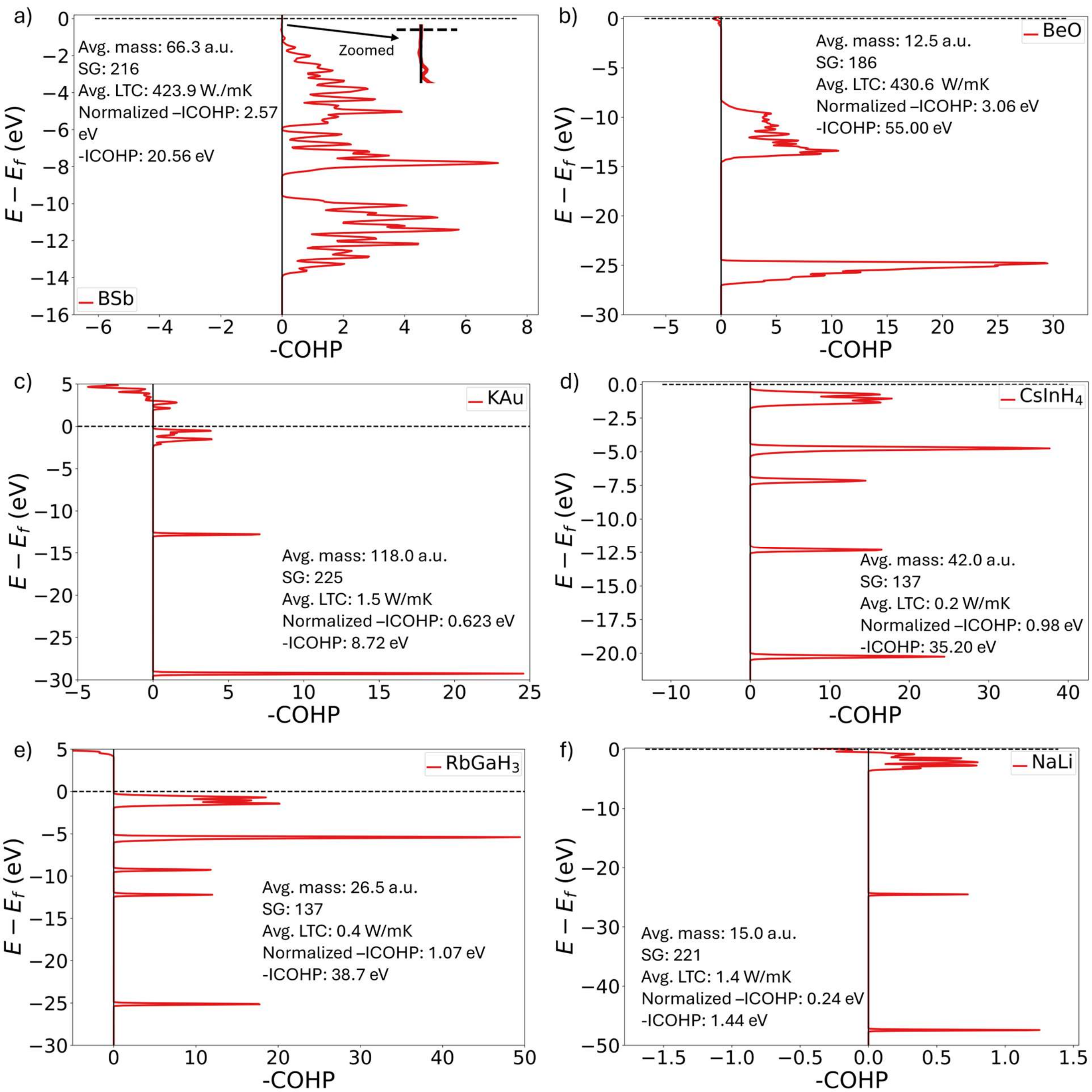

**Figure 6:** COHP curves of (a) BSb (OQMD ID 1218583), (b) BeO (OQMD ID 3167), (c) KAu (OQMD ID 1105148), (d) $CsInH_4$ (OQMD 1732883), and (e) $RbGaH_3$ (OQMD ID 1369521), (f) NaLi (OQMD 307185). The energy of -COHP curves are scaled such that the Fermi energy is zero. The dashed horizontal line represents the Fermi energy. -COHP curves of (a-b) are added to prove that high LTC materials can have antibonding states under Fermi energy while normalized -ICOHP is high. -COHP curves in (b-f) prove that antibonding states under Fermi energy are not necessary to have low LTC while normalized -ICOHP is low, contrary to Refs. [22, 78] that attributed low LTC to antibonding states.

In Figure 6, we further prove that our chemical bonding descriptors are more universal by the -COHP curves of some materials that Refs. [22, 52-56] failed to explain. The -COHP curve is shown instead of COBI because COHP had been used to explain LTC in previous studies and conclusions were drawn in those studies [22, 52-56], whereas COBI was not used before our study to explain LTC. Note that the -COHP curves in KAu, $CsInH_4$, and $RbGaH_3$ as shown in Figure 6c, 6d, and 6e have low normalized -ICOHP without any antibonding states under Fermi energy, and their LTCs are 1.5, 0.2, and 1.4 W/mK, respectively. Moreover, they contain the elements K, Cs, and Rb, respectively. Ref. [56] attributed the low LTC in materials containing K, Cs, or Rb to the existence of the antibonding states in COHP, and Refs. [22, 52, 55] justified the low LTC in their systems similarly. However, KAu, $CsInH_4$, and $RbGaH_3$ having low LTC do not have antibonding states which demonstrates that the claim in Refs. [22, 52-56] is not always the case, but the low normalized -ICOHP and normalized ICOBI are observed in those low LTC materials. The existence of the antibonding states reduces -ICOHP and consequently reduces normalized -ICOHP, but it is not a determining factor to judge whether such materials will certainly have low LTC. For example, the -COHP curves of BeO and BSb in Figure 6a and 6b, respectively, show antibonding states, but their LTCs are not low at all, namely, LTC of BeO and BSb are 423.9 and 430.6 W/mK, respectively. In Figure 6e, BSb has a higher average mass of 15 a.u. than NaLi, but the LTC of BSb (430.6 W/mK) is more than two orders of magnitude higher than that of NaLi (1.4 W/mK). This demonstrates that a high (low) average mass is also not enough to be solely used for screening low (high) LTC materials. We attribute the low LTC of NaLi to the weak chemical bonding strength between the constituent atoms of Na and Li, and such weak chemical bonding strength can be illustrated by the low normalized -ICOHP value of 0.24 eV, which is as low as many low LTC materials in Table S2.

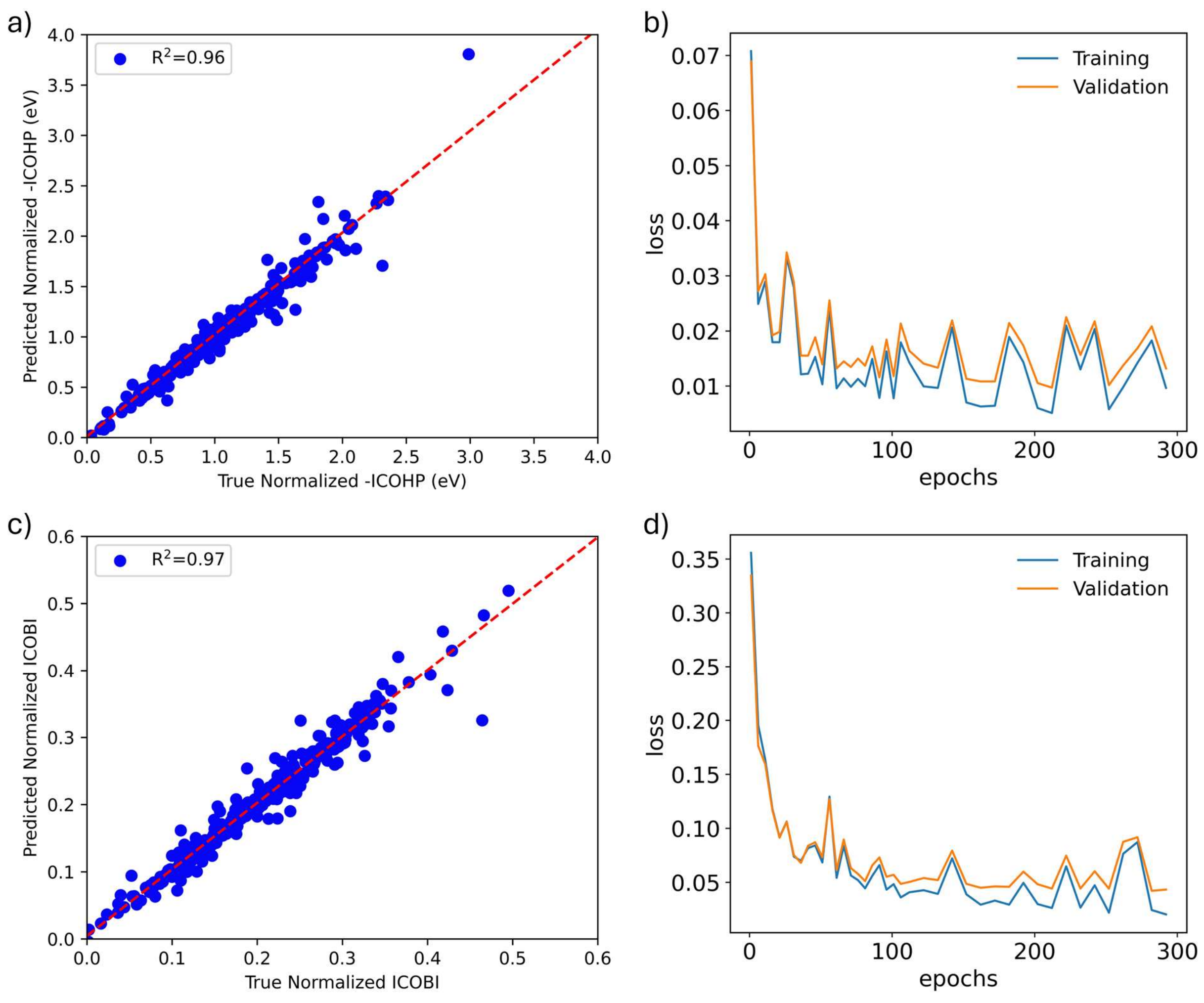

**Figure 7:** CATGNN model predictions vs. true values of the testing set for (a) normalized -ICOHP and (c) normalized ICOBI. MAE training and validation losses of our CATGNN model for (b) normalized -ICOHP and (d) normalized ICOBI.

**Table 2:** Comparison of $R^2$ and MAE of the four machine learning models trained for the two proposed chemical bonding descriptors, namely normalized -ICOHP and normalized ICOBI: (1) our newly developed CATGNN, (2) CGCNN, (3) Gradient Boosting, and (4) Random Forest.

| **Machine Learning Model** | **Normalized -ICOHP** | | **Normalized ICOBI** | |
|---|---|---|---|---|
| | **$R^2$ (%)** | **MAE (eV)** | **$R^2$ (%)** | **MAE** |
| CATGNN | 97.4 | 0.039 | 97.8 | 0.008 |
| CGCNN | 94.4 | 0.071 | 94.2 | 0.014 |
| Gradient Boosting | 82.2 | 0.155 | 88.3 | 0.020 |
| Random Forest | 0.06 | 0.288 | 79.8 | 0.028 |

### *d) CATGNN training results*

Based on the strong correlation between our newly defined chemical bonding descriptors and LTC in Figure 3a and 3b, it is intuitive to train ML models for such descriptors and then use the ML models to accelerate the search of new materials with extreme LTCs. To this end, we further trained four machine learning models for the two discovered chemical bonding descriptors, namely normalized -ICOHP and normalized ICOBI: (1) our newly developed CATGNN, (2) CGCNN, (3) Gradient Boosting, and (4) Random Forest. The training, validation, and testing dataset split is 75%, 15%, 10%, respectively. The dataset split of materials that contain specific elements can be shown in Figure S2 in Supplemental Information. AdamW optimizer was used in the model with a learning rate of 0.001. The performance parameters for each model are the coefficient of determination ($R^2$) and mean absolute error (MAE). The MAE is defined as $\frac{\sum_{i=0}^{N} |y_{prediction} - y_{true}|}{N}$, where $N$ is the total number of samples, and the subscript "prediction" and "true" means the predicted values by the ML models and true values calculated by SCF DFT, respectively. The performance of our CATGNN model is compared with Gradient Boosting and Random Forest trained utilizing the state-of-the-art magpie descriptors [57] and is also compared with crystal graph convolutional neural networks (CGCNN) [58], in terms of $R^2$ and MAE (see Table 2). Our CATGNN model outperforms the other three models, reflected by much higher $R^2$ and lower MAE. The CATGNN results are obtained through obtaining the weights of the best model from the entire 300 epochs trained. The best model is defined as the model whose validation loss is the lowest from the 300 epochs. The training and validation loss curves for the 300 epochs are shown in Figure 7a and 7b for normalized -ICOHP and normalized ICOBI, respectively. It seems that both models need approximately 200 epochs to reach convergence. The $R^2$ for the testing dataset is shown in Figure 7c and 7d for normalized -ICOHP and normalized ICOBI, respectively. The best model is then used to predict normalized -ICOHP and normalized ICOBI of 32,716, 11,069, and 156,138 non-zero bandgap materials from OQMD, ICSD, and materials cloud databases, respectively. In total, 900 materials with the lowest and highest normalized -ICOHP and normalized ICOBI are selected to screen for low and high LTC materials. To reduce DFT calculation time and to increase the chance to get truly dynamically stable materials, the phonon dispersions of the selected materials are first predicted using our newly trained CHGNet model from our own high precision SCF DFT dataset. Figure S4 in Supplemental Information shows phonon dispersions comparison of some selected materials predicted by our new CHGNet model and DFT. Then, after the dynamical stability is checked, i.e., imaginary frequencies are absent in the Brillouin zone, DFT calculations are then performed to obtain IFCs and then the LTCs. The success rate of CHGNet predicted dynamically stable materials is roughly 80%. After removing duplicate materials, LTC and average MSD of 106 stable materials with low normalized -ICOHP and normalized ICOBI are then calculated by DFT. Table 3 exhibits phonon and chemical bonding properties of the 25 lowest LTC materials. All 25 materials show ultralow LTC (less than 1 W/mK) and high average MSD, just as our proposed chemical bonding descriptors predict. The rest of the low LTC materials (all less than 5 W/mK) are presented in Table S2 in Supplemental Information. LTC and average MSD are also calculated for 15 materials with high normalized -ICOHP and normalized ICOBI after removing duplicate structures. Table 4 shows 15 stable materials with high normalized -ICOHP and normalized ICOBI. 13 materials show high LTC (higher than 100 W/mK) and low average MSD as expected from the correlation shown in Figure 3a and 3b. These results validate and reaffirm that our descriptors can be used for screening materials with extreme LTCs for various applications such as thermal insulation, thermoelectrics, and heat dissipation for electronic cooling.

**Table 3:** 25 stable and lowest lattice thermal conductivity (LTC) materials and our developed chemical bonding strength descriptors: normalized -ICOHP and normalized ICOBI. The LTCs reported here are averaged over three crystallographic directions. **Table S2** in Supplemental Information shows the rest of 81 low LTC materials.

| Database | ID | Reduced Formula | Space Group Number | LTC (W/mK) | Avg. MSD (Å²) | Normalized -ICOHP (eV) | Normalized ICOBI |
|---|---|---|---|---|---|---|---|
| OQMD | 1610433 | $Cs_4Na_2BiAs$ | 187 | 0.16 | 0.4240 | 0.5079 | 0.1328 |
| ICSD | 24366 | $BaI_2$ | 62 | 0.27 | 0.2628 | 0.3187 | 0.0598 |
| ICSD | 55138 | $Rb_2Te$ | 62 | 0.29 | 0.3286 | 0.4070 | 0.0916 |
| OQMD | 1610499 | $Rb_4Na_2SbAs$ | 187 | 0.32 | 0.2913 | 0.4423 | 0.1299 |
| OQMD | 1343293 | $K_2Te$ | 62 | 0.35 | 0.2921 | 0.3813 | 0.0939 |
| OQMD | 1610618 | $K_4Na_2BiAs$ | 187 | 0.36 | 0.2879 | 0.3820 | 0.1241 |
| OQMD | 1610485 | $KRb_3(NaSb)_2$ | 156 | 0.39 | 0.2726 | 0.4088 | 0.1287 |
| ICSD | 15706 | $BaBr_2$ | 62 | 0.41 | 0.2153 | 0.3634 | 0.0562 |
| OQMD | 1372624 | $Rb_2NaAs$ | 194 | 0.42 | 0.2871 | 0.4643 | 0.1299 |
| OQMD | 1610609 | KRbNaBi | 187 | 0.48 | 0.3023 | 0.3673 | 0.1244 |
| OQMD | 1610608 | $K_3Rb(NaBi)_2$ | 156 | 0.49 | 0.2920 | 0.3597 | 0.1239 |
| OQMD | 1610487 | KRbNaSb | 187 | 0.49 | 0.2723 | 0.3966 | 0.1275 |
| OQMD | 1617863 | $K_4Na_2SbAs$ | 187 | 0.50 | 0.2619 | 0.3963 | 0.1255 |
| OQMD | 1375291 | $Rb_2NaSb$ | 194 | 0.52 | 0.2748 | 0.4204 | 0.1297 |
| OQMD | 1610492 | KRbNaSb | 164 | 0.52 | 0.2657 | 0.4000 | 0.1277 |
| OQMD | 1617735 | $K_3Rb(NaSb)_2$ | 156 | 0.53 | 0.2625 | 0.3885 | 0.1266 |
| OQMD | 1369028 | $KRb_2GaSb_2$ | 72 | 0.56 | 0.2560 | 0.6114 | 0.1672 |
| ICSD | 72326 | KRbS | 62 | 0.59 | 0.2190 | 0.5151 | 0.0910 |
| OQMD | 61825 | $PbI_2$ | 164 | 0.61 | 1.0461 | 0.5074 | 0.1464 |
| ICSD | 300175 | $Na_5InS_4$ | 11 | 0.62 | 0.2549 | 0.7176 | 0.1478 |
| OQMD | 689141 | $K_2NaAs$ | 194 | 0.64 | 0.2475 | 0.4190 | 0.1252 |
| OQMD | 4412 | $PbI_2$ | 156 | 0.64 | 0.4401 | 0.5040 | 0.1465 |
| OQMD | 1610616 | $K_4Na_2BiSb$ | 187 | 0.65 | 0.2700 | 0.3664 | 0.1242 |
| OQMD | 1369307 | $K_3AlSb_2$ | 72 | 0.65 | 0.2462 | 0.7187 | 0.1976 |
| OQMD | 11304 | $PbI_2$ | 186 | 0.68 | 0.4470 | 0.5039 | 0.1465 |

**Table 4:** 15 stable and high LTC materials with corresponding to our developed chemical bonding strength descriptors: normalized -ICOHP and normalized ICOBI. Note that the structures from materials cloud have no pre-assigned ID. The formulas in parentheses are the full formulas for the materials.

| Database | ID | Reduced formula | Space Group # | LTC (W/mK) | Avg. MSD (Å²) | Normalized -ICOHP (eV) | Normalized ICOBI |
|---|---|---|---|---|---|---|---|
| OQMD | 599494 | C ($C_{42}$) | 166 | 2037.50 | 0.01097 | 4.8475 | 0.4808 |
| OQMD | 599491 | C ($C_{12}$) | 194 | 1889.03 | 0.00911 | 4.9582 | 0.5002 |
| OQMD | 637353 | C ($C_4$) | 139 | 1841.90 | 0.01022 | 4.8906 | 0.5019 |
| OQMD | 599492 | C ($C_{16}$) | 194 | 1755.40 | 0.00906 | 4.942 | 0.5002 |
| OQMD | 7497 | BN | 186 | 748.10 | 0.01135 | 4.6710 | 0.4533 |
| OQMD | 16167 | $BC_2N$ | 25 | 746.67 | 0.01086 | 4.6606 | 0.4777 |
| OQMD | 16166 | $BC_2N$ | 17 | 709.40 | 0.01077 | 4.6359 | 0.4724 |
| OQMD | 28581 | BP | 186 | 505.48 | 0.02160 | 3.2308 | 0.4968 |
| OQMD | 613825 | B ($B_{12}$) | 166 | 207.16 | 0.01992 | 3.9590 | 0.4578 |
| OQMD | 11716 | $B_6P$ | 166 | 184.07 | 0.01944 | 3.6795 | 0.4677 |
| OQMD | 18724 | $B_6As$ | 166 | 124.86 | 0.02085 | 3.6339 | 0.4675 |
| OQMD | 1474172 | AlN | 194 | 54.08 | 0.02189 | 3.5798 | 0.4280 |
| OQMD | 25648 | $SiB_3$ | 74 | 37.91 | 0.02624 | 3.3582 | 0.4641 |
| Materials Cloud | N/A | BN | 216 | 839.60 | 0.01140 | 4.8164 | 0.4658 |
| Materials Cloud | N/A | BP | 216 | 506.80 | 0.02140 | 3.2766 | 0.4988 |

To get further insight, we calculated more detailed statistics of element distribution in extreme LTC materials in Figure 8. Several elements have ultralow normalized -ICOHP and normalized ICOBI such as group 1 elements namely Na, K, Rb, and Cs, as reported previously [56]. Those elements tend to form some compounds that will have low LTC or high average MSD. Halogen elements, namely F, Cl, Br, and I, have high occurrences for ultralow normalized -ICOHP and normalized ICOBI and thus also have high chance to form materials with low LTC or high average MSD. On the other hand, elements such as Be, B, C, N, Al, and Si have higher averages of normalized -ICOHP and normalized ICOBI than many other elements in the periodic table. Those elements are constituent in materials with high LTC or low average MSD as shown in Table 4. It is also intriguing to observe high normalized ICOBI in some transition metals, such as Ti, V, Cr, Mn, Fe, Co, Nb, Mo, Ru, Hf, Ta, W, Re, Os, and Ir. Transition metals tend to have high melting temperatures due to forming strong covalent bonds with other unfilled d-shell valence electrons [51], and thus are also good candidate elements to constitute high LTC materials. Since COBI is a measure of covalency between atoms, it physically makes sense that many transition elements have higher normalized ICOBI than many other elements in the periodic table. These observations can be very helpful for designing novel desirable thermal materials with target LTC and MSD.

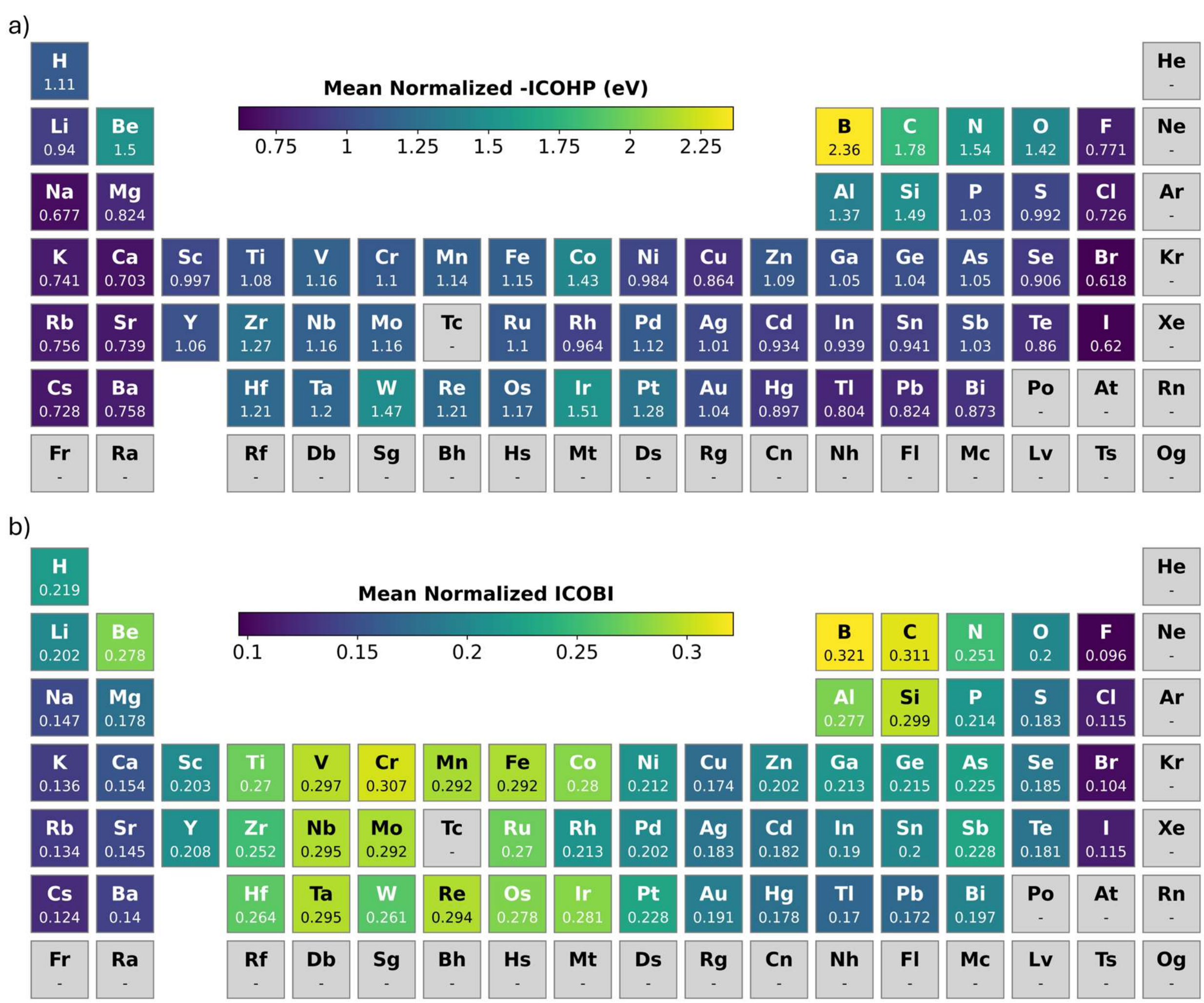

**Figure 8:** Periodic table element distribution color mapped with (a) normalized -ICOHP and (b) normalized ICOBI. The number below the element name in panel (a) and (b) is calculated by averaging normalized -ICOHP and ICOBI for all materials that contain that particular element, respectively.

### *e) Conclusions and Summary*

Designing new materials with extraordinary thermal transport properties needs unravelling atomic level hidden structure-property relationships and exceptional chemical bonding intuition. We developed two chemical bonding descriptors, namely normalized -ICOHP and normalized ICOBI, to correlate with LTC and the average MSD. Through comparing the normalized -ICOHP and average mass, we observe that for materials with the same or similar average mass the LTC increases as normalized -ICOHP increases, and vice versa. We also observed that the average MSD decreases as normalized -ICOHP increases and vice versa, without the average mass having any noticeable effect on the average MSD. The same trend was observed when comparing the normalized ICOBI and average mass with LTC and the average MSD. We compared normalized -ICOHP and normalized ICOBI of more than 4,500 materials with varying LTCs and observed that LTC is high (low) only when normalized -ICOHP and normalized ICOBI are high (low) *without* having to consider average mass as a descriptor. We also found that as the average MSD increases (decreases), normalized -ICOHP and normalized ICOBI decrease (increase). That indicates the inverse correlation between our developed chemical bonding descriptors with LTC and average MSD. Our new descriptors outperform the previous sole -ICOHP quantity by testing on a wide range of material classes and have much higher Pearson correlation with LTC and the average MSD than the traditional average mass. We trained our newly developed CATGNN model on those chemical bonding datasets and then predicted the normalized -ICOHP and normalized ICOBI of almost 200,000 structures from existing databases. We selected a total of 900 materials near the lower and upper bounds of normalized -ICOHP and normalized ICOBI to discover new materials with extreme LTCs. After initial dynamic stability screening by CHGNet model and confirmation by DFT, 13 stable materials with the high normalized -ICOHP and normalized ICOBI were found to have LTC higher than 100 W/mK, while 106 stable materials with the lowest normalized -ICOHP and normalized ICOBI turned out to possess a low LTC less than 5 W/mK with ~68% having less than 2 W/mK. These results not only accelerate the search for extreme LTCs but also shed light on electronic level descriptors to enhance our physical and chemical intuition in understanding phonon thermal transport in inorganic crystals. The workflow established here can be also implemented in the inverse design of novel materials with desirable extreme LTCs by quickly screening materials based on our intuitive chemical bonding principles.

## Methods

### *a) DFT calculations*

Primitive cells of all the structures are re-optimized using first-principles calculations performed by Vienna Ab-Initio Simulation package (VASP) [59-61]. Strict optimization convergence criteria for our calculations are $10^{-8}$ eV and $10^{-4}$ eV/Å for energy and atomic forces, respectively. The cell volume and shape along with the internal atomic positions are all allowed to change during the structure optimization. The generalized gradient approximation (GGA) of Pedrew-Burke-Ernzerhof (PBE) [62] is implemented to characterize the electrons exchange-correlation effects within projector augmented wave (PAW) pseudopotentials [63]. The kinetic energy cutoff for the plane-wave basis is set to be 520 eV to compute the electron charge density in all materials. The Monkhorst-Pack [64] wavevector (k)-mesh sampling is implemented in the DFT calculations, and for electrons the product of k-points along a specific crystallographic direction and corresponding lattice constant in Angstrom equal to 80 or above is used in the structure optimization. Regarding the phonon calculations, we generate 12 to 30 supercells with random displacement of 0.03 Å for all atoms in the supercells. The size of the supercell depends on the materials symmetry and the lattice constants. Generally speaking, the number of atoms in those supercells is between 80 and 300. The atomic forces were then calculated using self-consistent field (SCF) DFT with $10^{-6}$ eV being the energy convergence criterion and for electrons the product of k-points along a specific crystallographic direction of supercell and corresponding lattice constant in Angstrom is reduced to 60. We used compressive sensing lattice dynamics (CSLD) method [65-67] to fit the harmonic (2$^{nd}$ order) and anharmonic (3$^{rd}$ order) interatomic force constants (IFCs). The phonon dispersions are plotted to confirm the dynamic stability of the materials using Phonopy [68]. Afterwards, we calculate LTC using ShengBTE package with the 2$^{nd}$ and 3$^{rd}$ order IFCs as input [69]. The grid of phonon mesh is dense enough to ensure the total number of scattering channels exceeding $10^{8}$. Mean squared displacement (MSD) results at 300 K are obtained using Phonopy from the 2$^{nd}$ order IFC. We have compared the relevant phonon transport properties of some representative materials by our DFT calculations with other DFT studies and also experimental measurements in our recent works [34-35, 39]. The good agreement of these results show that the pseudopotentials we used are appropriate for generating high accuracy phonon data.

*b) Normalized -ICOHP and normalized ICOBI: Chemical Bonding Descriptors*

Calculation of our new descriptors depends on first calculating Crystal orbital Hamilton population (COHP) and crystal orbital bond index (COBI) performed using LOBSTER software (version 4.1.0) [70-75]. The calculations are performed after the SCF DFT calculations that provided the all-electron wavefunction in plane-wave basis to convert it into linear combinations of atomic orbitals (LCAO) basis. The LCAO basis was generated by projecting chemically intuitive orbitals for each species onto all-electron wavefunction with plane-wave basis from VASP. These basis functions and input files were generated with the help of pymatgen [76-77]. All the LOBSTER calculations are performed on the primitive cells of the materials, as also recommended by pymatgen. The charge spilling from converting the plane-wave basis to LCAO basis is 5% as recommended by LOBSTER and implemented in this work. As a result, the number of materials with successful COHP and COBI calculations decreased from 4,777 to 4,554 and 4,552, respectively.

COHP is defined as partitioning the electronic band structure in terms of the orbital-pair contribution by their Hamiltonian

$$H_{\mu\nu} = < \phi_{\mu} | \hat{H} | \phi_{\nu} > \quad (2)$$

where $H$ is the Hamiltonian, $\phi_{\mu}$ is orbital $\mu$, and $\phi_{\nu}$ is orbital $\nu$. Wavevector k-dependent LCAO basis at band $j$ has the following form $\phi_j(k,\ r) = c_{j\mu}(k)\phi_{\mu}(r) + c_{j\nu}(k)\phi_{\nu}(r)$ ..., where $c_{j\mu}$ and $c_{j\nu}$ are the coefficients for orbitals $\mu$ and $\nu$. The coefficients are used to construct the projected density matrix $P_{\mu\nu} = \sum_{j}^{MOs} f_j c_{j,\mu} c_{j,\nu}$. Energy-dependent COHP can be defined as

$$COHP_{\mu\nu}(E) = H_{\mu\nu} \sum_{j,k} Re(c_{\mu,jk}^{*} c_{\nu,jk}) \cdot \delta(\varepsilon_j(k) - E) \quad (3)$$

Crystal orbital bond index (COBI) quantifies covalent bonding in solid-state materials. As covalency increases, COBI increases and vice versa. Energy-dependent COBI can be expressed as

$$COBI_{\mu\nu}(E) = P_{\mu\nu} \sum_{j,k} Re(c_{\mu,jk}^{*} c_{\nu,jk}) \cdot \delta(\varepsilon_j(k) - E) \quad (4)$$

Integrated COHP and COBI (ICOHP and ICOBI) are obtained by integration up to the Fermi energy, which can be utilized as a convenient value to characterize chemical bonding.

$$ICOHP_{\mu\nu} = \int_{-\infty}^{\varepsilon_F} COHP_{\mu\nu}(E) dE \quad (5a)$$

$$ICOBI_{\mu\nu} = \int_{-\infty}^{\varepsilon_F} COBI_{\mu\nu}(E) dE \quad (5b)$$

As can be seen from COHP and COBI equations, their values can increase depending on the selected projected orbitals and the number of those projected orbitals as well, i.e., the more chemically intuitive projected orbitals there are, the higher the COHP and COBI values. In order to properly quantify those values, we propose the following definitions

$$normalized\ -ICOHP = \frac{\sum_{\mu}\sum_{\nu} ICOHP_{\mu\nu}}{\#\ of\ projected\ orbitals\ \ from\ all\ sites} \quad (6a)$$

$$normalized\ ICOBI = \frac{\sum_{\mu}\sum_{\nu} ICOBI_{\mu\nu}}{\#\ of\ projected\ orbitals\ from\ all\ sites} \qquad (6b)$$

In order to normalize ICOHP and ICOBI, they are divided by the number of chemically intuitive projected orbitals from all sites. Note that we use the negative value of ICOHP, because the bonding states in COHP under Fermi energy are negative. Therefore, the negative value of ICOHP is the chemical bonding strength descriptor for the material. The bonding states of COBI are positive under Fermi energy, and ICOBI is the chemical bonding descriptor for the material. We emphasize that such normalization is crucial as demonstrated by Table S1 in Supplemental Information and comparing Figures 2 – 3 with Figures S1 in Supplemental Information (see more details below). To reproduce the results of our work, we also added the POTCAR symbol and title used in our work in Table S3. We also include the basis set used for each element implemented in our LOBSTER calculations in Table S3 to reproduce this work. However, the above normalization definitions in Eqs. (6a) and (6b) are not strict. For example, we normalize by the number of atoms and show the definitions in Eqs. (S1a) and (S1b) along with Figure S5 which shows this new normalization definition results for LTC and average MSD. We would like to point out that using different functionals and pseudopotentials might affect the electronic structure of crystals, which might subsequently affect COHP and COBI results. However, LOBSTER only supports PAW type of pseudopotentials, so ultrasoft pseudopotentials do not work. LOBSTER also supports the PBE functional which is why in “lobsterin” file, “pbeVaspFit2015” is for the standard command “basisSet” [73].

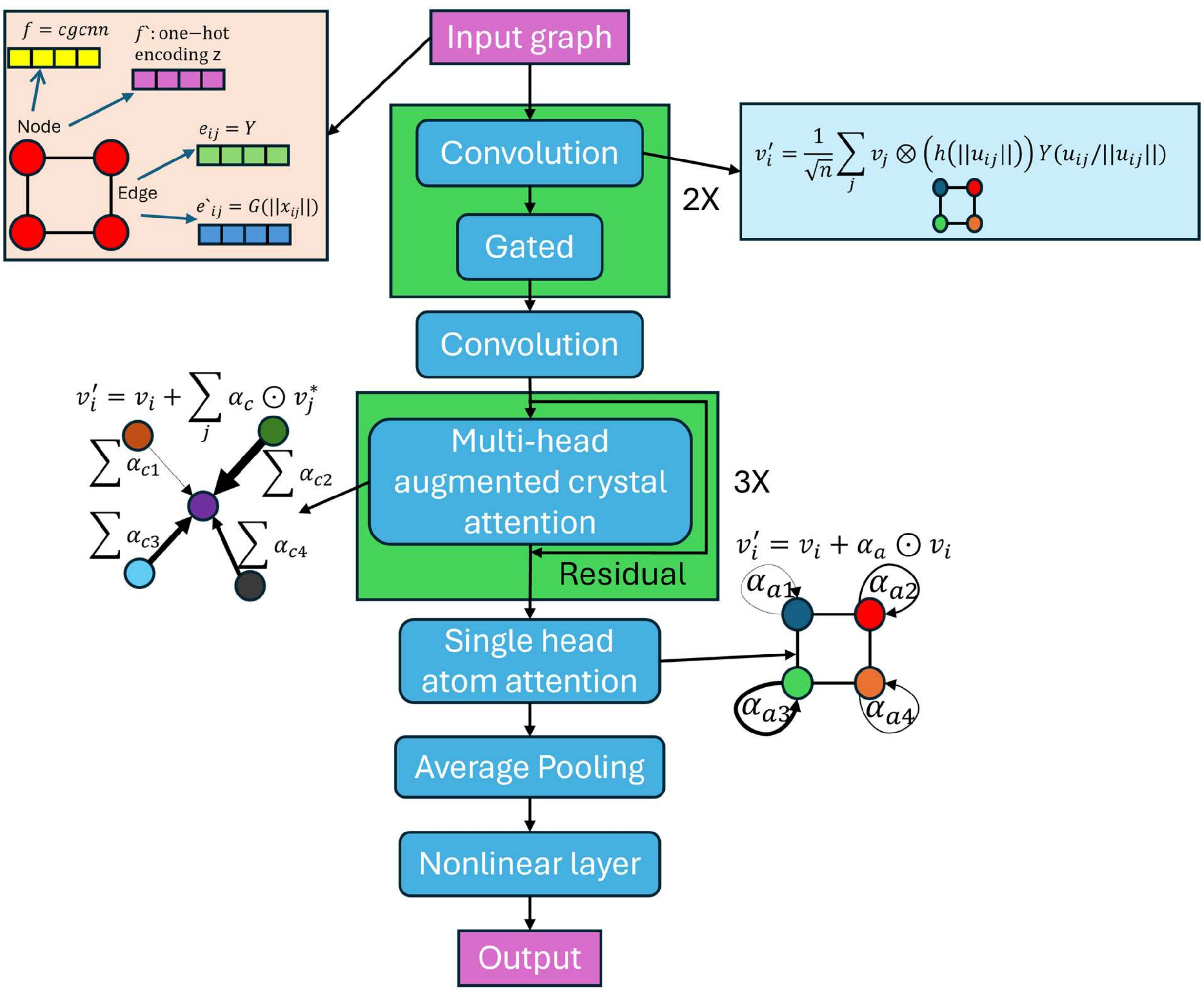

**Figure 9:** Schematic of architecture of our Crystal Attention Graph Neural Network (CATGNN) model.

### c) *CATGNN Architecture*

Our Crystal Attention Graph Neural Network (CATGNN) model is built utilizing Pytorch [78]. The general model architecture is shown in Figure 9. For each node in the graph, the node features and node attributes are represented by $f$ and $\grave{f}$, respectively. The node features ($f$) are the CGCNN features [58]. The node attributes ($\grave{f}$) are one-hot encoding atomic number feature vector. The edge features and edge attribute are represented by $e_{ij}$ and $\grave{e}_{ij}$, respectively. The edge features ($e_{ij}$) are spherical harmonics and the edge attributes ($\grave{e}_{ij}$) are gaussian expansion of the bond lengths. The convolutional layers are defined as

$$v_i' = \frac{1}{\sqrt{n}} \sum_j v_j \otimes \left(h(||x_{ij}||)\right) Y(x_{ij}/||x_{ij}||) \tag{7}$$

where n, $v_i'$, $v_j$, $x_{ij}$, $h$, and $Y$ are the number of neighbors, updated nodes feature vectors, input nodes feature vectors, relative displacement vector, multi-layer perceptron, and spherical harmonics, respectively. Those layers were directly imported from e3nn [79] with minor modifications to the layers to accurately consider periodic boundary conditions of crystals. The following layers named multi-head augmented crystal attention and single-head atom attention layers are developed in this model from scratch. The multi-head augmented crystal attention layer is defined as

$$v_i' = v_i + \sum_j \frac{\sum_h^{hs} softmax\left(g\left(BN\left(W_{att} \odot g\left(g(W_i \odot v_i^*) \oplus g\left(W_j \odot v_j^*\right)\right)\right)\right)\right) \odot v_j^*}{hs} = v_i + \sum_j \alpha_c \odot v_j^*$$

$$\alpha_c = softmax\left(g\left(BN\left(W_{att} \odot g\left(g(W_i \odot v_i^*) \oplus g\left(W_j \odot v_j^*\right)\right)\right)\right)\right) \tag{8}$$

where $g$ is the activation function which is softplus in this work. "+" indicates that two matrices are added. However, "$\oplus$" indicates that two matrices are concatenated instead of added. The number of attention heads symbolized by "$hs$" in $\alpha$ is 8. "$BN$" symbolizes a batch normalization layer. The feature vectors $v_i^*$ and $v_j^*$ in Eq. (8) are defined as

$$v_i^* = v_i \oplus (W_{i*} \odot u_{ij} + b_{i*}) \tag{9a}$$

$$v_j^* = v_j \oplus (W_{j*} \odot u_{ij} + b_{j*}) \tag{9b}$$

where $u_{ij}$ are the edge attributes represented by the Gaussian basis feature vector for the bond lengths between source node $i$ and destination node $j$. The Atom attention layer is defined as

$$v_i' = v_i + softmax\left(W_{atomAtt2}\left(BN\left(g\left(W_{atomAtt1}\left(v_i \oplus v_{iCgcnn}^*\right) + b_{atomAtt1}\right)\right)\right) + b_{atomAtt2}\right) \odot v_i$$

$$= v_i + \alpha_a \odot v_i$$

$$\alpha_a = softmax\left(W_{atomAtt2}\left(BN\left(g\left(W_{atomAtt}\ \left(v_i \oplus v_{iCgcnn}^*\right) + b_{atomAtt1}\right)\right)\right) + b_{atomAtt2}\right)$$

$$v_{iCgcnn}^* = g(W_{iCgcnn} \odot v_{iCgcnn} + b_{iCgcnn}) \tag{10}$$

## Acknowledgements

This work was supported in part by the NSF (award number 2110033, 2311202, 2320292) and SC EPSCoR/IDeA Program under NSF OIA-1655740 (23-GC01). R.R. acknowledges financial support by the Severo Ochoa Centres of Excellence Program under grant CEX2023-001263-S, and by the Generalitat de Catalunya under grant 2021 SGR 01519. Calculations were performed at the Centro de Supercomputación de Galicia (CESGA) within actions FI-2023-1-0003, FI-2023-2-0005, and FI-2024-1-0012 of the Red Española de Supercomputación (RES).

## Author Contributions

M.H. conveyed the idea and designed and supervised the study. M.A. developed the crystal attention graph neural network model, performed training and testing, and conducted the LOBSTER calculations. R.R. performed self-consistent field DFT calculations. M.A. prepared the draft of the manuscript. R.R., J.H., C.W. and M.H. revised the manuscript. All authors contributed to discussions and interpretation of results in the manuscript.

## Competing Interests

The Authors declare no Competing Financial or Non-Financial Interests.

## Code Availability

CATGNN code will be shared on a GitHub repository in https://github.com/Mofahdi/CATGNN. Other codes on how to obtain the chemical bonding descriptors (normalized -ICOHP and normalized ICOBI) will also be available in the GitHub repository as well.

## Data Availability

The 4,554 normalized -ICOHP and 4,552 normalized ICOBI datasets used for training the CATGNN model are provided in the Excel file named “supplementary data.xlsx”.

## References

[1] Tritt, T. M. Thermal Conductivity: Theory, Properties, and Applications; Springer International Publishing, 2005.
[2] ROWE, D. M. CRC Handbook of Thermoelectrics; CRC Press, 1995.
[3] Al-Fahdi, M.; Zhang, X.; Hu, M. Phonon Transport Anomaly in Metavalent Bonded Materials: Contradictory to the Conventional Theory. Journal of Materials Science, 56, 18534–18549, (2021).
[4] Chang, Z.; Ma, J.; Yuan, K.; Zheng, J.; Wei, B.; Al-Fahdi, M.; Gao, Y.; Zhang, X.; Shao, H.; Hu, M.; Tang, D. Zintl Phase Compounds Mg3Sb2−xBix (x = 0, 1, and 2) Monolayers: Electronic, Phonon and Thermoelectric Properties from Ab Initio Calculations. Frontiers in Mechanical Engineering, 8, (2022).
[5] Lin, S.; Yue, jincheng; Ren, W.; Shen, C.; Zhang, H. Strong Anharmonicity and Medium-Temperature Thermoelectric Efficiency in Antiperovskites CA3XN (x = P, as, Sb, Bi) Compounds. Journal of Materials Chemistry A (2024).
[6] Tian, X.; Chen, M. Descriptor Selection for Predicting Interfacial Thermal Resistance by Machine Learning Methods. Scientific Reports, 11, 1, (2021).
[7] Al-Fahdi, M.; Hu, M. High Throughput Substrate Screening for Interfacial Thermal Management of β-Ga2o3 by Deep Convolutional Neural Network. Journal of Applied Physics, 135, (2024).
[8] Wu, J.; Zhou, E.; Huang, A.; Zhang, H.; Hu, M.; Qin, G. Deep-Potential Enabled Multiscale Simulation of Gallium Nitride Devices on Boron Arsenide Cooling Substrates. Nature Communications, 15, (2024).
[9] Wang, H.; Wei, D.; Duan, J.; Qin, Z.; Qin, G.; Yao, Y.; Hu, M. The Exceptionally High Thermal Conductivity after 'Alloying' Two-Dimensional Gallium Nitride (Gan) and Aluminum Nitride (Aln). Nanotechnology, 32, 135401, (2021).
[10] Yue, J.; Liu, Y.; Ren, W.; Lin, S.; Shen, C.; Kumar Singh, H.; Cui, T.; Tadano, T.; Zhang, H. Role of Atypical Temperature-Responsive Lattice Thermal Transport on the Thermoelectric Properties of Antiperovskites Mg3xn (x = P, as, Sb, Bi). Materials Today Physics, 41, 101340, (2024).
[11] Zhu, H.; Mao, J.; Feng, Z.; Sun, J.; Zhu, Q.; Liu, Z.; Singh, D. J.; Wang, Y.; Ren, Z. Understanding the Asymmetrical Thermoelectric Performance for Discovering Promising Thermoelectric Materials. Science Advances, 5, (2019).
[12] Tadano, T.; Gohda, Y.; Tsuneyuki, S. Impact of Rattlers on Thermal Conductivity of a Thermoelectric Clathrate: A First-Principles Study. Physical Review Letters, 114, (2015).
[13] Lin, H.; Tan, G.; Shen, J.; Hao, S.; Wu, L.; Calta, N.; Malliakas, C.; Wang, S.; Uher, C.; Wolverton, C.; Kanatzidis, M. G. Concerted Rattling in CSAG5Te3 Leading to Ultralow Thermal Conductivity and High Thermoelectric Performance. Angewandte Chemie, 128, 11603–11608, (2016).
[14] Zhou, Y.; Yang, J.-Y.; Cheng, L.; Hu, M. Strong Anharmonic Phonon Scattering Induced Giant Reduction of Thermal Conductivity in PBTE Nanotwin Boundary. Physical Review B, 97, (2018).
[15] Sarkar, D.; Ghosh, T.; Roychowdhury, S.; Arora, R.; Sajan, S.; Sheet, G.; Waghmare, U. V.; Biswas, K. Ferroelectric Instability Induced Ultralow Thermal Conductivity and High Thermoelectric Performance in Rhombohedral p-Type Gese Crystal. Journal of the American Chemical Society, 142, 12237–12244, (2020).
[16] Skoug, E. J.; Morelli, D. T. Role of Lone-Pair Electrons in Producing Minimum Thermal Conductivity in Nitrogen-Group Chalcogenide Compounds. Physical Review Letters, 107, (2011).
[17] Qin, G.; Wang, H.; Qin, Z.; Hu, M. Activated Lone-Pair Electrons Lead to Low Lattice Thermal Conductivity: A Case Study of Boron Arsenide. SSRN Electronic Journal (2021).

[18] Carnevali, V.; Mukherjee, S.; Voneshen, D. J.; Maji, K.; Guilmeau, E.; Powell, A. V.; Vaqueiro, P.; Fornari, M. Lone Pair Rotation and Bond Heterogeneity Leading to Ultralow Thermal Conductivity in Aikinite. Journal of the American Chemical Society, 145, 9313–9325, (2023).
[19] Z. Chang, K. Yuan, J. Li, Z. Sun, J. Zheng, M. Al-Fahdi, Y. Gao, B. Wei, X. Zhang, M. Hu, and D. Tang, "Anomalous thermal conductivity induced by high dispersive optical phonons in rubidium and cesium halides," EES 16, 30–39 (2022).
[20] Keyes, R. W. High-Temperature Thermal Conductivity of Insulating Crystals: Relationship to the Melting Point. Physical Review, 115, 564–567, (1959).
[21] Wang, F. Q.; Hu, M.; Wang, Q. Ultrahigh Thermal Conductivity of Carbon Allotropes with Correlations with the Scaled Pugh Ratio. Journal of Materials Chemistry A, 7, 6259–6266, (2019).
[22] He, J.; Xia, Y.; Lin, W.; Pal, K.; Zhu, Y.; Kanatzidis, M. G.; Wolverton, C. Accelerated Discovery and Design of Ultralow Lattice Thermal Conductivity Materials Using Chemical Bonding Principles. Advanced Functional Materials, 32, (2021).
[23] Qin, G.; Huang, A.; Liu, Y.; Wang, H.; Qin, Z.; Jiang, X.; Zhao, J.; Hu, J.; Hu, M. High-Throughput Computational Evaluation of Lattice Thermal Conductivity Using an Optimized Slack Model. Materials Advances, 3, 6826–6830, (2022).
[24] Qin, G.; Qin, Z.; Wang, H.; Hu, M. Anomalously Temperature-Dependent Thermal Conductivity of Monolayer Gan with Large Deviations from the Traditional 1/T Law. Physical Review B, 95, (2017).
[25] Yuan, K.; Zhang, X.; Tang, D.; Hu, M. Anomalous Pressure Effect on the Thermal Conductivity of Zno, Gan, and AlN from First-Principles Calculations. Physical Review B, 98, (2018).
[26] Yue, S.-Y.; Ouyang, T.; Hu, M. Diameter Dependence of Lattice Thermal Conductivity of Single-Walled Carbon Nanotubes: Study from Ab Initio. Scientific Reports 2015, 5 (1).
[27] Qin, G.; Hu, M. Accelerating Evaluation of Converged Lattice Thermal Conductivity. npj Computational Materials, 4, (2018).
[28] Gorai, P.; Stevanović, V.; Toberer, E. S. Computationally Guided Discovery of Thermoelectric Materials. Nature Reviews Materials, 2, (2017).
[29] Seko, A.; Togo, A.; Hayashi, H.; Tsuda, K.; Chaput, L.; Tanaka, I. Prediction of Low-Thermal-Conductivity Compounds with First-Principles Anharmonic Lattice-Dynamics Calculations and Bayesian Optimization. Physical Review Letters, 115, (2015).
[30] Al-Fahdi, M.; Ouyang, T.; Hu, M. High-Throughput Computation of Novel Ternary B–C–N Structures and Carbon Allotropes with Electronic-Level Insights into Superhard Materials from Machine Learning. Journal of Materials Chemistry A, 9, 27596–27614, (2021).
[31] Al-Fahdi, M.; Rodriguez, A.; Ouyang, T.; Hu, M. High-Throughput Computation of New Carbon Allotropes with Diverse Hybridization and Ultrahigh Hardness. Crystals, 11, 783, (2021).
[32] Fan, Q.; Min, G.; Liu, L.; Zhao, Y.; Yu, X.; Yun, S. Accelerate the Design of New Superhard Carbon Allotropes in PCA21 Space Group: High-Throughput Screening and Machine Learning Strategies. Diamond and Related Materials, 143, 110928, (2024).
[33] Ojih, J.; Al-Fahdi, M.; Rodriguez, A. D.; Choudhary, K.; Hu, M. Efficiently Searching Extreme Mechanical Properties via Boundless Objective-Free Exploration and Minimal First-Principles Calculations. npj Computational Materials, 8, (2022).
[34] Rodriguez, A.; Lin, C.; Yang, H.; Al-Fahdi, M.; Shen, C.; Choudhary, K.; Zhao, Y.; Hu, J.; Cao, B.; Zhang, H.; Hu, M. Million-Scale Data Integrated Deep Neural Network for Phonon Properties of Heuslers Spanning the Periodic Table. npj Computational Materials, 9, (2023).
[35] Rodriguez, A.; Lin, C.; Shen, C.; Yuan, K.; Al-Fahdi, M.; Zhang, X.; Zhang, H.; Hu, M. Unlocking Phonon Properties of a Large and Diverse Set of Cubic Crystals by Indirect Bottom-up Machine Learning Approach. Communications Materials, 4, (2023).

[36] Ojih, J.; Al-Fahdi, M.; Yao, Y.; Hu, J.; Hu, M. Graph Theory and Graph Neural Network Assisted High-Throughput Crystal Structure Prediction and Screening for Energy Conversion and Storage. Journal of Materials Chemistry A, 12, 8502–8515, (2024).
[37] Chen, L.; Tran, H.; Batra, R.; Kim, C.; Ramprasad, R. Machine Learning Models for the Lattice Thermal Conductivity Prediction of Inorganic Materials. *Computational Materials Science*, *170*, 109155, (**2019**).
[38] Zhu, T.; He, R.; Gong, S.; Xie, T.; Gorai, P.; Nielsch, K.; Grossman, J. C. Charting Lattice Thermal Conductivity for Inorganic Crystals and Discovering Rare Earth Chalcogenides for Thermoelectrics. Energy & Environmental Science, 14, 3559–3566, (2021).
[39] Ojih, J.; Onyekpe, U.; Rodriguez, A.; Hu, J.; Peng, C.; Hu, M. Machine Learning Accelerated Discovery of Promising Thermal Energy Storage Materials with High Heat Capacity. *ACS Applied Materials & Interfaces*, *14*, 43277–43289, (**2022**).
[40] Al-Fahdi, M.; Yuan, K.; Yao, Y.; Rurali, R.; Hu, M. High-Throughput Thermoelectric Materials Screening by Deep Convolutional Neural Network with Fused Orbital Field Matrix and Composition Descriptors. Applied Physics Reviews, 11, (2024).
[41] Long, T.; Fortunato, N. M.; Zhang, Y.; Gutfleisch, O.; Zhang, H. An Accelerating Approach of Designing Ferromagnetic Materials via Machine Learning Modeling of Magnetic Ground State and Curie Temperature. Materials Research Letters, 9, 169–174, (2021).
[42] Lightstone, J. P.; Chen, L.; Kim, C.; Batra, R.; Ramprasad, R. Refractive Index Prediction Models for Polymers Using Machine Learning. Journal of Applied Physics, 127, (2020).
[43] Sun, J.; Zhang, C.; Yang, Z.; Shen, Y.; Hu, M.; Wang, Q. Four-Phonon Scattering Effect and Two-Channel Thermal Transport in Two-Dimensional Paraelectric Snse. ACS Applied Materials & Interfaces, 14, 11493–11499, (2022).
[44] Ouyang, Y.; Yu, C.; He, J.; Jiang, P.; Ren, W.; Chen, J. Accurate Description of High-Order Phonon Anharmonicity and Lattice Thermal Conductivity from Molecular Dynamics Simulations with Machine Learning Potential. Physical Review B, 105, (2022).
[45] Zhao, Y.; Al-Fahdi, M.; Hu, M.; Siriwardane, E. M.; Song, Y.; Nasiri, A.; Hu, J. High-throughput Discovery of Novel Cubic Crystal Materials Using Deep Generative Neural Networks. Advanced Science, 8, (2021).
[46] Zhao, Y.; Siriwardane, E. M.; Wu, Z.; Fu, N.; Al-Fahdi, M.; Hu, M.; Hu, J. Physics Guided Deep Learning for Generative Design of Crystal Materials with Symmetry Constraints. npj Computational Materials, 9, (2023).
[47] Saal, J. E.; Kirklin, S.; Aykol, M.; Meredig, B.; Wolverton, C. Materials Design and Discovery with High-Throughput Density Functional Theory: The Open Quantum Materials Database (OQMD). JOM, 65, 1501–1509, (2013).
[48] Bergerhoff, G.; Hundt, R.; Sievers, R.; Brown, I. D. The Inorganic Crystal Structure Data Base. Journal of Chemical Information and Computer Sciences, 23, 66–69, (1983).
[49] Belsky, A.; Hellenbrandt, M.; Karen, V. L.; Luksch, P. New Developments in the Inorganic Crystal Structure Database (ICSD): Accessibility in Support of Materials Research and Design. Acta Crystallographica Section B Structural Science, 58, 364–369, (2002).
[50] Talirz, L.; et al. Materials Cloud, a platform for open computational science. Sci. Data, 7, 299, (2020).
[51] Vedernikov, M. V. The Thermoelectric Powers of Transition Metals at High Temperature. Advances in Physics, 18, 337–370, (1969).
[52] Ding, J.; Lanigan-Atkins, T.; Calderón-Cueva, M.; Banerjee, A.; Abernathy, D. L.; Said, A.; Zevalkink, A.; Delaire, O. Soft Anharmonic Phonons and Ultralow Thermal Conductivity in $Mg_3(Sb, Bi)_2$ Thermoelectrics. Science Advances, 7, (2021).
[53] Jaffe, J. E.; Zunger, A. Theory of the Band-Gap Anomaly in $ABC_2$ Chalcopyrite Semiconductors. Physical Review B, 29, 1882–1906, (1984).
[54] Wei, S.-H.; Zunger, A. Role of Metal d States in II-VI Semiconductors. Physical Review B, 37, 8958–8981, (1988).

[55] Cai, W.; He, J.; Li, H.; Zhang, R.; Zhang, D.; Chung, D. Y.; Bhowmick, T.; Wolverton, C.; Kanatzidis, M. G.; Deemyad, S. Pressure-Induced Ferroelectric-like Transition Creates a Polar Metal in Defect Antiperovskites Hg3Te2x2 (X = Cl, Br). Nature Communications 2021, 12 (1).
[56] Yuan, J.; Chen, Y.; Liao, B. Lattice Dynamics and Thermal Transport in Semiconductors with Anti-Bonding Valence Bands. Journal of the American Chemical Society, 145, 18506–18515, (2023).
[57] Ward, L.; Agrawal, A.; Choudhary, A.; Wolverton, C. A General-Purpose Machine Learning Framework for Predicting Properties of Inorganic Materials. npj Computational Materials, 2, (2016).
[58] Xie, T.; Grossman, J. C. Crystal Graph Convolutional Neural Networks for an Accurate and Interpretable Prediction of Material Properties. Physical Review Letters, 120, (2018).
[59] Kresse, G.; Furthmüller, J. Efficiency of AB-Initio Total Energy Calculations for Metals and Semiconductors Using a Plane-Wave Basis Set. *Comput. Mater. Sci.*, *6*, 15–50, (**1996**).
[60] Kresse, G.; Furthmüller, J. Efficient Iterative Schemes for *Ab Initio* Total-Energy Calculations Using a Plane-Wave Basis Set. *Phys. Rev. B*, *54*, 11169–11186, (**1996**).
[61] Kresse, G.; Joubert, D. From Ultrasoft Pseudopotentials to the Projector Augmented-Wave Method. *Phys. Rev. B*, *59*, 1758–1775, (**1999**).
[62] Perdew, J. P.; Burke, K.; Ernzerhof, M. Generalized Gradient Approximation Made Simple. *Phys. Rev. Lett.*, *77*, 3865–3868, (**1996**).
[63] Blöchl, P. E. Projector Augmented-Wave Method. *Physical Review B*, *50*, 17953–17979, (**1994**).
[64] Monkhorst, H. J.; Pack, J. D. Special Points for Brillouin-Zone Integrations. *Phys. Rev. B*, *13*, 5188–5192, (**1976**).
[65] Fei Zhou, Weston Nielson, Yi Xia, and Vidvuds Ozoliņš. Lattice Anharmonicity and Thermal Conductivity from Compressive Sensing of First-Principles Calculations. *Physical Review Letters* 113, 185501 (2014).
[66] Fei Zhou, Weston Nielson, Yi Xia, and Vidvuds Ozoliņš. Compressive sensing lattice dynamics. I. General formalism. *Physical Review B* 100, 184308 (2019).
[67] Fei Zhou, Babak Sadigh, Daniel Åberg, Yi Xia, and Vidvuds Ozoliņš. Compressive sensing lattice dynamics. II. Efficient phonon calculations and long-range interactions. *Physical Review B* 100, 184309 (2019).
[68] Togo, A.; Tanaka, I. First Principles Phonon Calculations in Materials Science. *Scr. Mater.*, *108*, 1–5, (**2015**).
[69] Li, W.; Carrete, J.; A. Katcho, N.; Mingo, N. Shengbte: A Solver of the Boltzmann Transport Equation for Phonons. *Comput. Phys. Commun.*, *185*, 1747–1758, (**2014**).
[70] Dronskowski, R.; Bloechl, P. E. Crystal Orbital Hamilton Populations (COHP): Energy-Resolved Visualization of Chemical Bonding in Solids Based on Density-Functional Calculations. The Journal of Physical Chemistry, 97, 8617–8624, (1993).
[71] Deringer, V. L.; Tchougréeff, A. L.; Dronskowski, R. Crystal Orbital Hamilton Population (COHP) Analysis as Projected from Plane-Wave Basis Sets. The Journal of Physical Chemistry A, 115, 5461–5466, (2011).
[72] Maintz, S.; Deringer, V. L.; Tchougréeff, A. L.; Dronskowski, R. Analytic Projection from Plane-Wave and Paw Wavefunctions and Application to Chemical-Bonding Analysis in Solids. Journal of Computational Chemistry, 34, 2557–2567, (2013).
[73] Maintz, S.; Deringer, V. L.; Tchougréeff, A. L.; Dronskowski, R. Lobster: A Tool to Extract Chemical Bonding from Plane-wave Based DFT. Journal of Computational Chemistry, 37, 1030–1035, (2016).
[74] Nelson, R.; Ertural, C.; George, J.; Deringer, V. L.; Hautier, G.; Dronskowski, R. Lobster: Local Orbital Projections, Atomic Charges, and Chemical-bonding Analysis from Projector-augmented-wave-based Density-functional Theory. Journal of Computational Chemistry, 41, 1931–1940, (2020).

[75] Müller, P. C.; Ertural, C.; Hempelmann, J.; Dronskowski, R. Crystal Orbital Bond Index: Covalent Bond Orders in Solids. The Journal of Physical Chemistry C, 125, 7959–7970, (2021).
[76] Ong, S. P.; Richards, W. D.; Jain, A.; Hautier, G.; Kocher, M.; Cholia, S.; Gunter, D.; Chevrier, V. L.; Persson, K. A.; Ceder, G. Python Materials Genomics (Pymatgen): A Robust, Open-Source Python Library for Materials Analysis. Computational Materials Science, 68, 314–319, (2013).
[77] George, J.; Petretto, G.; Naik, A.; Esters, M.; Jackson, A. J.; Nelson, R.; Dronskowski, R.; Rignanese, G.; Hautier, G. Automated Bonding Analysis with Crystal Orbital Hamilton Populations. ChemPlusChem, 87, (2022).
[78] Ketkar, N.; Moolayil, J. Automatic Differentiation in Deep Learning. Deep Learning with Python, 133–145, (2021).
[79] Geiger, M; Smidt, T.; e3nn: Euclidean Neural Networks, arXiv:2207.09453

*Supplementary Information for*

# Accelerated discovery of extreme lattice thermal conductivity by crystal graph attention networks and chemical bonding

Mohammed Al-Fahdi,[1] Riccardo Rurali,[2] Jianjun Hu,[3] Christopher Wolverton,[4] and Ming Hu[1,*]
[1]Department of Mechanical Engineering, University of South Carolina, Columbia, South Carolina 29208, USA
[2]Institut de Ciència de Materials de Barcelona, ICMAB–CSIC, Campus UAB, 08193 Bellaterra, Spain
[3]Department of Computer Science and Engineering, University of South Carolina, Columbia, South Carolina 29208, USA
[4]Department of Materials Science and Engineering, Northwestern University, Evanston, IL 60201, USA

## Additional notes on normalization for -ICOHP and ICOBI

As seen from Eqs. (2) – (5) in the main text, the ICOHP and ICOBI are computed from the projected orbitals on all sites to quantify their bonding and antibonding interactions to grasp bonding strength between the atoms in a material. It can be simply deduced that the more atoms exist in the primitive cell of a material, the more projected orbitals there are, and the higher -ICOHP and ICOBI values are. An appropriate way to normalize such value is to divide the number of projected orbitals from all the sites in the primitive cell to properly and accurately quantify the bond strength. It is worth reiterating that LOBSTER calculations are done on the primitive cells of the materials, not the conventional cells as executed by pymatgen. If -ICOHP and ICOBI are good bonding strength descriptors without normalization then $CsZrAgTe_3$, $RbHfCuSe_3$, and MgCdSO should have higher LTC than BAs, BN, BSb, and AlN since they have higher -ICOHP and ICOBI as shown in Table S1. However, the normalized -ICOHP and normalized ICOBI are higher in materials with high LTC such as BAs, BN, BSb, and AlN than materials with low LTC such as BaO, BaLiSb, $CsZrAgTe_3$, $RbHfCuSe_3$, and MgCdSO. Furthermore, the importance of normalization is not just seen between materials with high and low LTC, but it is also seen in the materials with the same reduced formula but with different primitive cell formula. For example, BAs with two different phases (i.e., cubic phase with space group number of 216 and hexagonal phase with space group number of 186) have the following primitive cell formula: BAs in the cubic phase and $B_2As_2$ in the hexagonal phase. The projected orbitals of B atom are 2s and 2p, and the projected orbitals of As atom are 4s, and 4p. Since the hexagonal phase has twice more atoms and therefore twice more projected orbitals than the cubic phase, it is not surprising that the hexagonal phase of BAs has higher -ICOHP and ICOBI (-ICOHP and ICOBI are approximately twice as high in the hexagonal phase than the cubic one). However, upon looking at the normalized -ICOHP and normalized ICOBI, the values are approximately the same (i.e., the normalized -ICOHP values are 3.032 eV and 3.020 eV for hexagonal and cubic phases, respectively, and the normalized -ICOBI values are 0.493 and 0.495 for hexagonal and cubic phases, respectively). The difference in normalized -ICOHP and normalized ICOBI can be attributed to the difference in the different number of nearest neighbors, bond lengths, local environment in each site due to the change in phase which caused the difference in the bonding interactions in both phases. The same observation of -ICOHP and

* Author to whom all correspondence should be addressed. E-Mail: hu@sc.edu (M.H.)

ICOBI being roughly a multiple in phases where the primitive cell number of atoms is a multiple of the reduced formula is also observed in other materials with high and low LTC such as AlN, BaO, BaLiSb, BN, and BSb as seen from Table S1. After realizing all the above statements, we claim our descriptors are properly normalized and can be used universally for a wide range of materials classes as can be seen from Table S1, Figures S1-S2, and Figures 2 – 3 in the main text.

In Figure S1 a), lots of the red dots (i.e., materials with high LTC) exist around the middle or middle to high part of the plot where -ICOHP is average or average-to-high. A vast number of materials with average LTC represented by the green color have higher -ICOHP than most of the materials with high LTC that are represented by the red color. However, when the normalized -ICOHP is compared with average mass and LTC in Figure 2 a) in the main text, materials with high LTC in red tend to have high normalized -ICOHP and low average mass. Moreover, the -ICOHP decrease does not in general clearly and necessarily mean that LTC decreases at the same average mass in Figure S1 a). Generally speaking, if the bond strength is high in materials, the materials could have high LTC because bond strength is directly proportional to phonon group velocity and LTC is directly proportional to phonon group velocity from Eq. (1) in the main text. The fact that many materials in green with medium LTC in Figure S1 a) have higher -ICOHP than many materials in red with high LTC proves that -ICOHP by itself is not a sufficiently good general bonding strength descriptor for LTC. On the other hand, the trend of high LTC having high normalized -ICOHP (i.e., the bond strength descriptor) and LTC decreasing as normalized -ICOHP decreases for the same mean atomic mass is clearly seen from Figure 2 a) in the main text. It can also be observed from Figure S1 b) that several materials with medium and medium-to-high average MSD in green, yellow, and orange have higher -ICOHP than many materials with low average MSD in dark and light blue across a wide range of average atomic masses. Moreover, materials with medium average MSD exist in the area under the low average MSD materials where -ICOHP are lower than the blue dots. However, if the bonding strength (-ICOHP) is a universally sufficient general descriptor for all the materials to capture bonding strength, then the materials with low average MSD in blue should have higher -ICOHP than the green ones with medium average MSD. As it is known, atoms with strong bonding strength should have deeper potential, and those atoms should not be displaced much from their equilibrium positions during thermal motion. The fact that many materials with medium average MSD in green above the low average MSD materials in blue demonstrates that -ICOHP descriptor is not sufficiently good universal descriptor to represent the general trend. Figure 2 b) in the main text shows a much better trend for the average MSD since it clearly shows that generally speaking the materials with high (low) average MSD have low (high) normalized -ICOHP. Furthermore, the colors change gradually from red to blue according to the color bar colors based on their average MSD values from high normalized -ICOHP to low normalized -ICOHP. Moreover, Figure S1 visually analyzes ICOBI vs. mean atomic mass with c) LTC and d) average MSD. As mentioned in the main text, ICOBI is a measure of bonds covalency. It is also known that covalent bonds are stronger than ionic bonds [1]. In Figure S1 c), a tremendously large number of materials with medium LTC in green and some blue dots that represent materials with low LTC have higher ICOBI than red dots with high LTC. That demonstrates that the ICOBI descriptor is not a sufficiently good bonding strength descriptor since bond strength is directly proportional to phonon group velocity which is also directly related to LTC. Therefore, the materials with higher bond strength would have higher LTC in general. Upon looking at Figure 2 c) the normalized ICOBI is a better descriptor for chemical bonding strength. In Figure S1 d), materials with medium average MSD represented by green from the color bar have higher ICOBI than many materials with low average MSD in light and dark blue colors. If ICOBI is a sufficiently good bonding strength descriptor for all materials classes, then not many materials with average MSD in green dots will appear above the blue dots that represent the materials with low average MSD. Low average MSD materials should have higher bonding strength since the interatomic potential in such materials should be deeper. When Figure S1d) is compared with Figure 2 d), it can clearly be seen that the average MSD trend makes more sense physically in Figure 2 d) because the average MSD gradually decreases based on the color bar from low to high bonding strength descriptor which is the normalized ICOBI.

In summary, according to Table S1, Figure S1, and Figures 2-3, normalized -ICOHP and normalized ICOBI are better chemical bonding strength descriptors compared to the direct values (unnormalized) of -ICOHP and ICOBI.

In Figure S2, the training/validation/testing sets fractional split, which is 75%/15%/10%, looks similar among most of the elements. That ensures that the elements-based training/validation/testing fractional splits are properly balanced. Figure S3 shows the element count for materials in the entire dataset.

In Figure S4, the phonon dispersion plots of DFT and our newly trained CHGNet model from scratch are compared with each other. It can be observed that the phonon dispersion plots from both DFT and our newly trained CHGNet model are approximately the same except that a systematic underestimation of frequencies from our trained CHGNet model clearly exists. However, our CHGNet model is still good enough to screen positive/ imaginary phonon frequencies and thus accelerate the discovery of dynamically stable materials.

Here we normalize the number of atoms to demonstrate that the normalization definitions in Eqs. (6a) and (6b) are not unique in showing good trends with LTC and average MSD. Eqs. (S1a) and (S1b) show those two new normalization definitions:

$$normalized\ -ICOHP\ (\#\ of\ atoms) = \frac{\sum_{\mu}\sum_{\nu} ICOHP_{\mu\nu}}{\#\ of atoms} \tag{S1a}$$

$$normalized\ ICOBI\ (\#\ of\ atoms) = \frac{\sum_{\mu}\sum_{\nu} ICOBI_{\mu\nu}}{\#\ of\ atoms} \tag{S1b}$$

Those two normalization definitions are analyzed with LTC and average MSD as shown in Figure S5. The plots show similar trends as Figures 2-3 in the main text.

**Table S1:** selected materials with OQMD ID, primitive cell formula, reduced formula, and space group (SG), their chemical bonding information (i.e., -ICOHP, normalized -ICOHP, ICOBI, and normalized ICOBI), and materials properties (i.e., LTC and the average MSD).

| OQMD ID | Reduced formula | Space group number | -ICOHP (eV) | Normalized -ICOHP (eV) | ICOBI | Normalized ICOBI | LTC (W/mk) | Avg. MSD (Å²) |
|---|---|---|---|---|---|---|---|---|
| 1440386 | AlN | 186 | 50.188 | 3.137 | 6.300 | 0.394 | 319.5 | 0.0231 |
| 1218324 | AlN | 216 | 26.780 | 3.347 | 3.402 | 0.425 | 275.0 | 0.0242 |
| 1105129 | BaO | 225 | 9.376 | 1.042 | 1.203 | 0.134 | 3.5 | 0.0945 |
| 22045 | BaO | 194 | 16.493 | 0.916 | 2.047 | 0.114 | 3.8 | 0.0902 |
| 1277969 | BAs | 186 | 48.505 | 3.032 | 7.891 | 0.493 | 1908.9 | 0.0263 |
| 8235 | BAs | 216 | 24.160 | 3.020 | 3.961 | 0.495 | 2370.4 | 0.0267 |
| 1489473 | BaLiSb | 62 | 24.875 | 0.565 | 6.688 | 0.152 | 2.8 | 0.1407 |
| 1455903 | BaLiSb | 189 | 17.192 | 0.521 | 4.986 | 0.151 | 1.1 | 0.1728 |
| 1455061 | BaLiSb | 194 | 12.601 | 0.573 | 3.707 | 0.169 | 5.6 | 0.1359 |
| 1236324 | BN | 216 | 38.296 | 4.787 | 3.729 | 0.466 | 844.3 | 0.0116 |
| 7497 | BN | 186 | 75.340 | 4.709 | 7.254 | 0.453 | 748.1 | 0.0115 |
| 1277988 | BSb | 186 | 41.083 | 2.568 | 7.858 | 0.491 | 375.1 | 0.0349 |
| 1218583 | BSb | 216 | 20.558 | 2.570 | 3.966 | 0.496 | 423.9 | 0.0358 |
| 1049981 | MgCdSO | 61 | 158.844 | 0.946 | 27.290 | 0.162 | 1.8 | 0.0965 |
| 1357607 | $CsZrAgTe_3$ | 62 | 143.826 | 0.999 | 32.140 | 0.223 | 0.5 | 0.1655 |
| 1357736 | $RbHfCuSe_3$ | 62 | 114.641 | 0.819 | 25.071 | 0.179 | 1.2 | 0.1172 |

**Table S2:** 81 selected stable materials with low LTC and high MSD and our developed chemical bonding strength descriptors: normalized -ICOHP and normalized ICOBI. The lattice thermal conductivities (LTCs) reported here are averaged over three crystallographic directions.

| **Database** | **Id** | **Reduced formula** | **Space group #** | **LTC (W/mk)** | **Avg. MSD (Å²)** | **Normalized -ICOHP (eV)** | **Normalized ICOBI** |
|---|---|---|---|---|---|---|---|
| OQMD | 1554232 | $Na_2Mg_2BiP$ | 99 | 0.75 | 0.2374 | 0.4914 | 0.1440 |
| OQMD | 11976 | $Na_6MnS_4$ | 186 | 0.78 | 0.1823 | 0.5737 | 0.1348 |
| OQMD | 1376579 | BaNaAs | 189 | 0.82 | 0.1254 | 0.4660 | 0.1309 |
| ICSD | 20020 | KI | 194 | 0.82 | 0.3203 | 0.3675 | 0.0679 |
| ICSD | 422273 | $K_2SrCdSb_2$ | 26 | 0.83 | 0.1672 | 0.6457 | 0.1663 |
| OQMD | 1369362 | $K_2NaSb$ | 194 | 0.84 | 0.2569 | 0.3798 | 0.1255 |
| ICSD | 41322 | CsNaSe | 129 | 0.92 | 0.1951 | 0.5341 | 0.1053 |
| OQMD | 1338889 | $Sr_7Ge_3$ | 186 | 0.93 | 0.1850 | 0.2229 | 0.0833 |
| OQMD | 1474283 | $K_2Te$ | 123 | 0.99 | 0.2980 | 0.3691 | 0.0950 |
| ICSD | 35458 | BaBrCl | 62 | 0.99 | 0.2552 | 0.3645 | 0.0542 |
| ICSD | 67276 | KNaTe | 62 | 1.03 | 0.2047 | 0.4077 | 0.1054 |
| OQMD | 24724 | $Na_3Sr_3GaP_4$ | 186 | 1.04 | 0.1228 | 0.6765 | 0.1630 |
| ICSD | 107569 | CsNaTe | 129 | 1.08 | 0.2262 | 0.4530 | 0.1058 |
| ICSD | 92771 | $Na_2S$ | 62 | 1.09 | 0.1714 | 0.4914 | 0.1116 |
| OQMD | 1561683 | $Ba_2TlCuSb_2$ | 156 | 1.10 | 0.1352 | 0.5757 | 0.1570 |
| OQMD | 1505450 | $Ca_2ZnSbP$ | 36 | 1.11 | 0.0987 | 0.7883 | 0.1852 |
| OQMD | 1554236 | $Na_2Mg_2BiAs$ | 99 | 1.11 | 0.1864 | 0.4797 | 0.1440 |
| OQMD | 1482262 | BaNaAs | 62 | 1.13 | 0.1294 | 0.4772 | 0.1327 |
| OQMD | 1505419 | Ca2CdBiAs | 36 | 1.18 | 0.1292 | 0.7106 | 0.1810 |
| OQMD | 1552765 | $KNa(MgBi)_2$ | 99 | 1.29 | 0.1972 | 0.4233 | 0.1370 |
| ICSD | 67278 | KNaSe | 62 | 1.40 | 0.1713 | 0.4565 | 0.1016 |
| OQMD | 1371419 | $K2NaGaBi_2$ | 72 | 1.46 | 0.2370 | 0.5980 | 0.1704 |
| OQMD | 1371775 | $Rb_2NaTlAs_2$ | 72 | 1.46 | 0.1910 | 0.6699 | 0.1706 |
| OQMD | 1554237 | $Na_2Mg_2SbAs$ | 99 | 1.49 | 0.1695 | 0.4979 | 0.1460 |
| OQMD | 1554435 | $RbNa(MgSb)_2$ | 99 | 1.50 | 0.1861 | 0.4522 | 0.1377 |
| ICSD | 409178 | RbCaSb | 129 | 1.51 | 0.1998 | 0.3848 | 0.1238 |
| OQMD | 1552767 | $RbNa(MgBi)_2$ | 99 | 1.52 | 0.2082 | 0.4197 | 0.1352 |
| OQMD | 1366391 | $Na_2MgGe$ | 194 | 1.56 | 0.1703 | 0.3506 | 0.1263 |
| OQMD | 1284285 | BaSrSi | 62 | 1.58 | 0.1632 | 0.4171 | 0.1292 |
| OQMD | 1553617 | $Rb_2CaMgP_2$ | 115 | 1.58 | 0.1493 | 0.5138 | 0.1273 |
| ICSD | 62658 | KNaS | 62 | 1.59 | 0.1525 | 0.4904 | 0.1000 |
| OQMD | 1374584 | $K_2NaTlAs_2$ | 72 | 1.60 | 0.1954 | 0.6589 | 0.1709 |
| OQMD | 1553454 | $K_2CaCdAs_2$ | 115 | 1.61 | 0.1451 | 0.6788 | 0.1677 |
| OQMD | 24717 | BaNaP | 189 | 1.64 | 0.1185 | 0.4909 | 0.1325 |
| OQMD | 1371513 | NaCaSb | 62 | 1.65 | 0.1216 | 0.3979 | 0.1324 |
| OQMD | 1366462 | $Cs_2NaGaSb_2$ | 72 | 1.71 | 0.2046 | 0.6600 | 0.1759 |
| OQMD | 1372533 | $Rb_2NaGaBi_2$ | 72 | 1.74 | 0.2313 | 0.6054 | 0.1708 |
| OQMD | 1753875 | $K_2NaTlP_2$ | 72 | 1.80 | 0.1801 | 0.6987 | 0.1730 |
| OQMD | 1374152 | $K_2LiInSb_2$ | 72 | 1.83 | 0.1834 | 0.6896 | 0.1866 |
| ICSD | 52681 | BaCaGe | 62 | 1.84 | 0.1233 | 0.2978 | 0.0981 |
| OQMD | 1554374 | $RbNa(MgAs)_2$ | 99 | 1.85 | 0.1535 | 0.5097 | 0.1377 |
| OQMD | 1553966 | $K_2CaMgP_2$ | 115 | 1.87 | 0.1471 | 0.4826 | 0.1276 |
| OQMD | 1385208 | NaCaAs | 189 | 1.92 | 0.1022 | 0.4411 | 0.1330 |
| OQMD | 1554235 | $Na_2Mg_2BiSb$ | 99 | 1.93 | 0.1797 | 0.4584 | 0.1452 |
| OQMD | 1553665 | $KNa(MgP)_2$ | 99 | 1.94 | 0.1400 | 0.5341 | 0.1386 |
| OQMD | 1377160 | NaMgBi | 129 | 1.98 | 0.1894 | 0.4434 | 0.1441 |
| ICSD | 172006 | SrCaGe | 62 | 1.99 | 0.1091 | 0.2891 | 0.1016 |
| OQMD | 1554384 | $CsNa(MgAs)_2$ | 99 | 2.00 | 0.1652 | 0.5446 | 0.1395 |
| OQMD | 1554378 | $KNa(MgAs)_2$ | 99 | 2.04 | 0.1501 | 0.5056 | 0.1378 |
| OQMD | 1742597 | KMgAs | 194 | 2.07 | 0.1688 | 0.5010 | 0.1530 |
| OQMD | 1505601 | $Ca_2CdAsP$ | 36 | 2.10 | 0.0934 | 0.7896 | 0.1834 |
| OQMD | 1552758 | $K_2Mg_2BiAs$ | 99 | 2.12 | 0.1442 | 0.4538 | 0.1300 |
| OQMD | 1017904 | $NaLi_2Bi$ | 139 | 2.18 | 0.1823 | 0.6740 | 0.1896 |
| OQMD | 26328 | NaSrP | 189 | 2.21 | 0.0975 | 0.4805 | 0.1345 |
| OQMD | 1473988 | $K_2Te$ | 225 | 2.22 | 0.2426 | 0.3902 | 0.1009 |
| ICSD | 42455 | $Ca_2Ge$ | 62 | 2.22 | 0.1138 | 0.2848 | 0.1000 |
| OQMD | 24739 | NaSrAs | 189 | 2.22 | 0.1057 | 0.4543 | 0.1324 |
| OQMD | 1501614 | $Ba_3Ca_3(As_2P)_2$ | 189 | 2.28 | 0.0860 | 0.5871 | 0.1435 |
| ICSD | 166534 | $Ca_2AsI$ | 166 | 2.33 | 0.1257 | 0.4051 | 0.1025 |
| OQMD | 1381978 | $Na_2LiSb$ | 194 | 2.33 | 0.1809 | 0.5291 | 0.1714 |
| ICSD | 172007 | SrCaSn | 62 | 2.36 | 0.1151 | 0.2658 | 0.0993 |
| OQMD | 1370558 | $Rb_2NaGaSb_2$ | 72 | 2.38 | 0.1996 | 0.6457 | 0.1749 |
| OQMD | 7598 | NaMgSb | 129 | 2.46 | 0.1690 | 0.4754 | 0.1464 |
| OQMD | 1745662 | NaCaP | 189 | 2.47 | 0.0932 | 0.4589 | 0.1339 |
| OQMD | 1369685 | $Ca_2MgAs_2$ | 36 | 2.56 | 0.0926 | 0.4664 | 0.1285 |
| ICSD | 172602 | $Sr_2IN$ | 166 | 2.50 | 0.0955 | 0.5273 | 0.1144 |
| OQMD | 1554373 | $KRb(MgAs)_2$ | 99 | 2.68 | 0.1241 | 0.5148 | 0.1311 |
| OQMD | 1552760 | $K_2Mg_2SbAs$ | 99 | 2.86 | 0.1340 | 0.4754 | 0.1328 |
| ICSD | 6068 | $Ca_2PI$ | 166 | 2.93 | 0.1120 | 0.4144 | 0.1039 |
| OQMD | 1553662 | $KRb(MgP)_2$ | 99 | 2.97 | 0.1129 | 0.5504 | 0.1329 |
| OQMD | 1553664 | $CsK(MgP)_2$ | 99 | 3.15 | 0.1297 | 0.5753 | 0.1339 |
| OQMD | 1374151 | $K_2LiGaSb_2$ | 72 | 3.40 | 0.1734 | 0.7228 | 0.1889 |
| ICSD | 172600 | $Sr_2BrN$ | 166 | 3.46 | 0.0910 | 0.5602 | 0.1095 |
| ICSD | 42458 | SrMgGe | 62 | 3.57 | 0.0879 | 0.3263 | 0.1043 |
| ICSD | 65216 | $Ca_2IN$ | 194 | 3.66 | 0.0735 | 0.5090 | 0.1175 |
| OQMD | 1376631 | CsMgP | 129 | 3.76 | 0.1418 | 0.6164 | 0.1363 |
| OQMD | 30639 | KMgP | 129 | 4.26 | 0.1080 | 0.5364 | 0.1319 |
| OQMD | 28055 | KMgAs | 129 | 4.34 | 0.1193 | 0.5064 | 0.1316 |
| ICSD | 41959 | CaTe | 194 | 4.87 | 0.0962 | 0.4273 | 0.1058 |
| ICSD | 153105 | $Ca_2BrN$ | 166 | 4.97 | 0.0731 | 0.5327 | 0.1084 |
| OQMD | 1743569 | $Ca_3Mg_2(CuP_2)_2$ | 164 | 5.18 | 0.0671 | 0.5087 | 0.1308 |

**Table S3:** POTCAR symbol and title along with the implemented LOBSTER basis for each element. The elements are ordered alphabetically.

| Element | POTCAR Symbol | POTCAR Title | LOBSTER Basis |
|---|---|---|---|
| Ag | Ag | PAW_PBE Ag 02Apr2005 | 4d 5p 5s |
| Al | Al | PAW_PBE Al 04Jan2001 | 3p 3s |
| As | As | PAW_PBE As 22Sep2009 | 4p 4s |
| Au | Au | PAW_PBE Au 04Oct2007 | 5d 6p 6s |
| B | B | PAW_PBE B 06Sep2000 | 2p 2s |
| Ba | Ba_sv | PAW_PBE Ba_sv 06Sep2000 | 5p 5s 6s |
| Be | Be_sv | PAW_PBE Be_sv 06Sep2000 | 1s 2p 2s |
| Bi | Bi | PAW_PBE Bi 08Apr2002 | 6p 6s |
| Br | Br | PAW_PBE Br 06Sep2000 | 4p 4s |
| C | C | PAW_PBE C 08Apr2002 | 2p 2s |
| Ca | Ca_sv | PAW_PBE Ca_sv 06Sep2000 | 3p 3s 4s |
| Cd | Cd | PAW_PBE Cd 06Sep2000 | 4d 5p 5s |
| Cl | Cl | PAW_PBE Cl 06Sep2000 | 3p 3s |
| Co | Co | PAW_PBE Co 02Aug2007 | 3d 4p 4s |
| Cr | Cr_pv | PAW_PBE Cr_pv 02Aug2007 | 3d 3p 4s |
| Cs | Cs_sv | PAW_PBE Cs_sv 08Apr2002 | 5p 5s 6s |
| Cu | Cu_pv | PAW_PBE Cu_pv 06Sep2000 | 3d 3p 4s |
| F | F | PAW_PBE F 08Apr2002 | 2p 2s |
| Fe | Fe_pv | PAW_PBE Fe_pv 02Aug2007 | 3d 3p 4s |
| Ga | Ga_d | PAW_PBE Ga_d 06Jul2010 | 3d 4p 4s |
| Ge | Ge_d | PAW_PBE Ge_d 03Jul2007 | 3d 4p 4s |
| H | H | PAW_PBE H 15Jun2001 | 1s |
| Hf | Hf_pv | PAW_PBE Hf_pv 06Sep2000 | 5d 5p 6s |
| Hg | Hg | PAW_PBE Hg 06Sep2000 | 5d 6p 6s |
| I | I | PAW_PBE I 08Apr2002 | 5p 5s |
| In | In_d | PAW_PBE In_d 06Sep2000 | 4d 5p 5s |
| Ir | Ir | PAW_PBE Ir 06Sep2000 | 5d 6p 6s |
| K | K_sv | PAW_PBE K_sv 06Sep2000 | 3p 3s 4s |

| Li | Li_sv | PAW_PBE Li_sv 10Sep2004 | 1s 2s 2p |
|---|---|---|---|
| Mg | Mg_pv | PAW_PBE Mg_pv 13Apr2007 | 2p 3s |
| Mn | Mn_pv | PAW_PBE Mn_pv 02Aug2007 | 3d 3p 4s |
| Mo | Mo_pv | PAW_PBE Mo_pv 04Feb2005 | 4d 4p 5s |
| N | N | PAW_PBE N 08Apr2002 | 2p 2s |
| Na | Na_pv | PAW_PBE Na_pv 19Sep2006 | 2p 3s |
| Nb | Nb_pv | PAW_PBE Nb_pv 08Apr2002 | 4d 4p 5s |
| Ni | Ni_pv | PAW_PBE Ni_pv 06Sep2000 | 3d 3p 4s |
| O | O | PAW_PBE O 08Apr2002 | 2p 2s |
| Os | Os_pv | PAW_PBE Os_pv 20Jan2003 | 5d 5p 6s |
| P | P | PAW_PBE P 06Sep2000 | 3p 3s |
| Pb | Pb_d | PAW_PBE Pb_d 06Sep2000 | 5d 6p 6s |
| Pd | Pd | PAW_PBE Pd 04Jan2005 | 4d 5p 5s |
| Pt | Pt | PAW_PBE Pt 04Feb2005 | 5d 6p 6s |
| Rb | Rb_sv | PAW_PBE Rb_sv 06Sep2000 | 4p 4s 5s |
| Re | Re_pv | PAW_PBE Re_pv 06Sep2000 | 5d 5p 6s |
| Rh | Rh_pv | PAW_PBE Rh_pv 25Jan2005 | 4d 4p 5s |
| Ru | Ru_pv | PAW_PBE Ru_pv 28Jan2005 | 4d 4p 5s |
| S | S | PAW_PBE S 06Sep2000 | 3p 3s |
| Sb | Sb | PAW_PBE Sb 06Sep2000 | 5p 5s |
| Sc | Sc_sv | PAW_PBE Sc_sv 07Sep2000 | 3d 3p 3s 4s |
| Se | Se | PAW_PBE Se 06Sep2000 | 4p 4s |
| Si | Si | PAW_PBE Si 05Jan2001 | 3p 3s |
| Sn | Sn_d | PAW_PBE Sn_d 06Sep2000 | 4d 5p 5s |
| Sr | Sr_sv | PAW_PBE Sr_sv 07Sep2000 | 4p 4s 5s |
| Ta | Ta_pv | PAW_PBE Ta_pv 07Sep2000 | 5d 5p 6s |
| Te | Te | PAW_PBE Te 08Apr2002 | 5p 5s |
| Ti | Ti_pv | PAW_PBE Ti_pv 07Sep2000 | 3d 3p 4s |

| Tl | Tl_d | PAW_PBE Tl_d 06Sep2000 | 5d 6p 6s |
|---|---|---|---|
| V | V_pv | PAW_PBE V_pv 07Sep2000 | 3d 3p 4s |
| W | W_sv | PAW_PBE W_sv 04Sep2015 | 5d 5p 5s 6s |
| Y | Y_sv | PAW_PBE Y_sv 25May2007 | 4d 4p 4s 5s |
| Zn | Zn | PAW_PBE Zn 06Sep2000 | 3d 4p 4s |
| Zr | Zr_sv | PAW_PBE Zr_sv 04Jan2005 | 4d 4p 4s 5s |

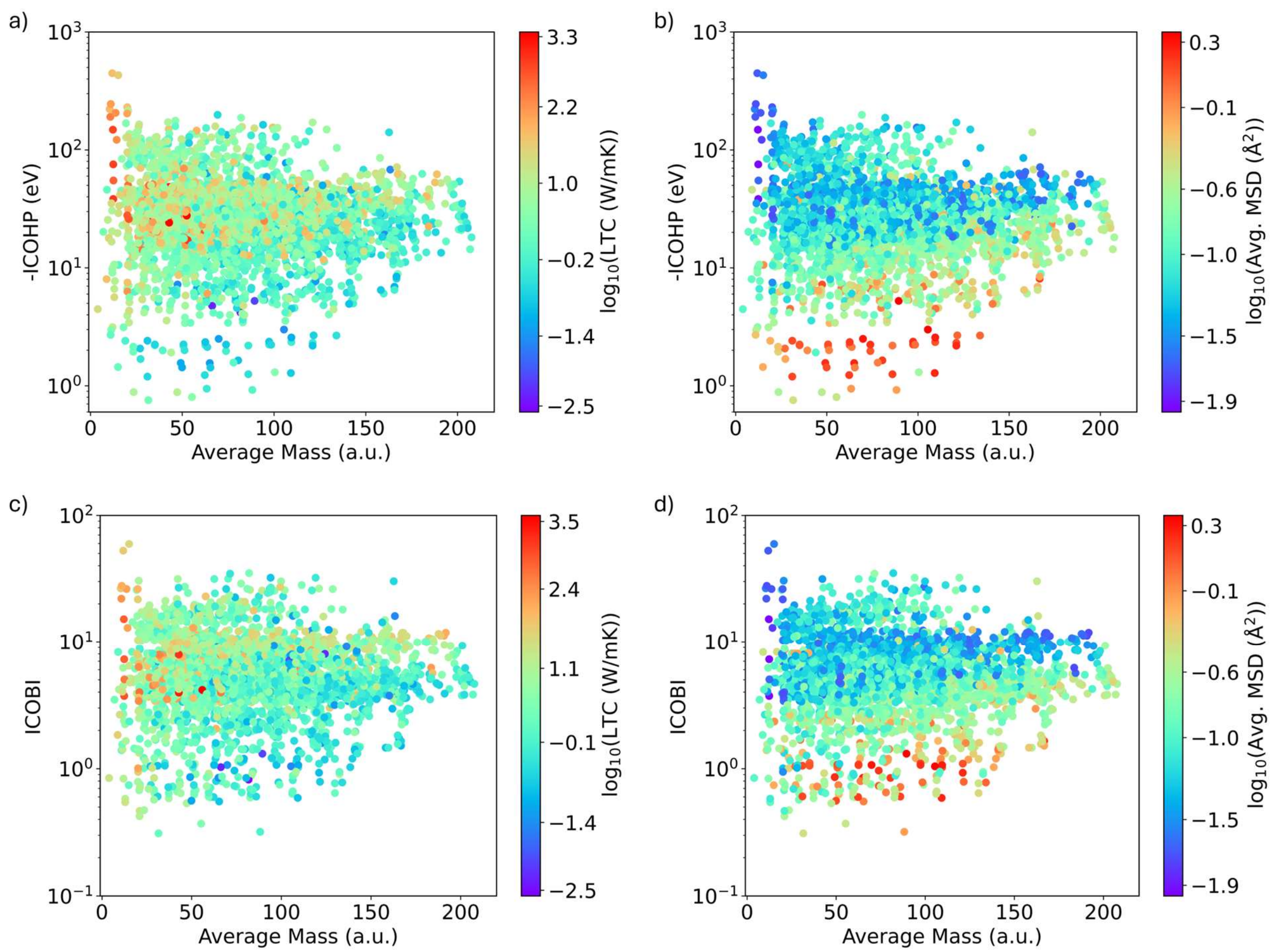

**Figure S1:** -ICOHP vs. average mass color mapped with (a) $\log_{10}$(LTC) and (b) $\log_{10}$(Avg. MSD). ICOBI vs. average mass color mapped with (c) $\log_{10}$(LTC) and (d) $\log_{10}$(Avg. MSD).

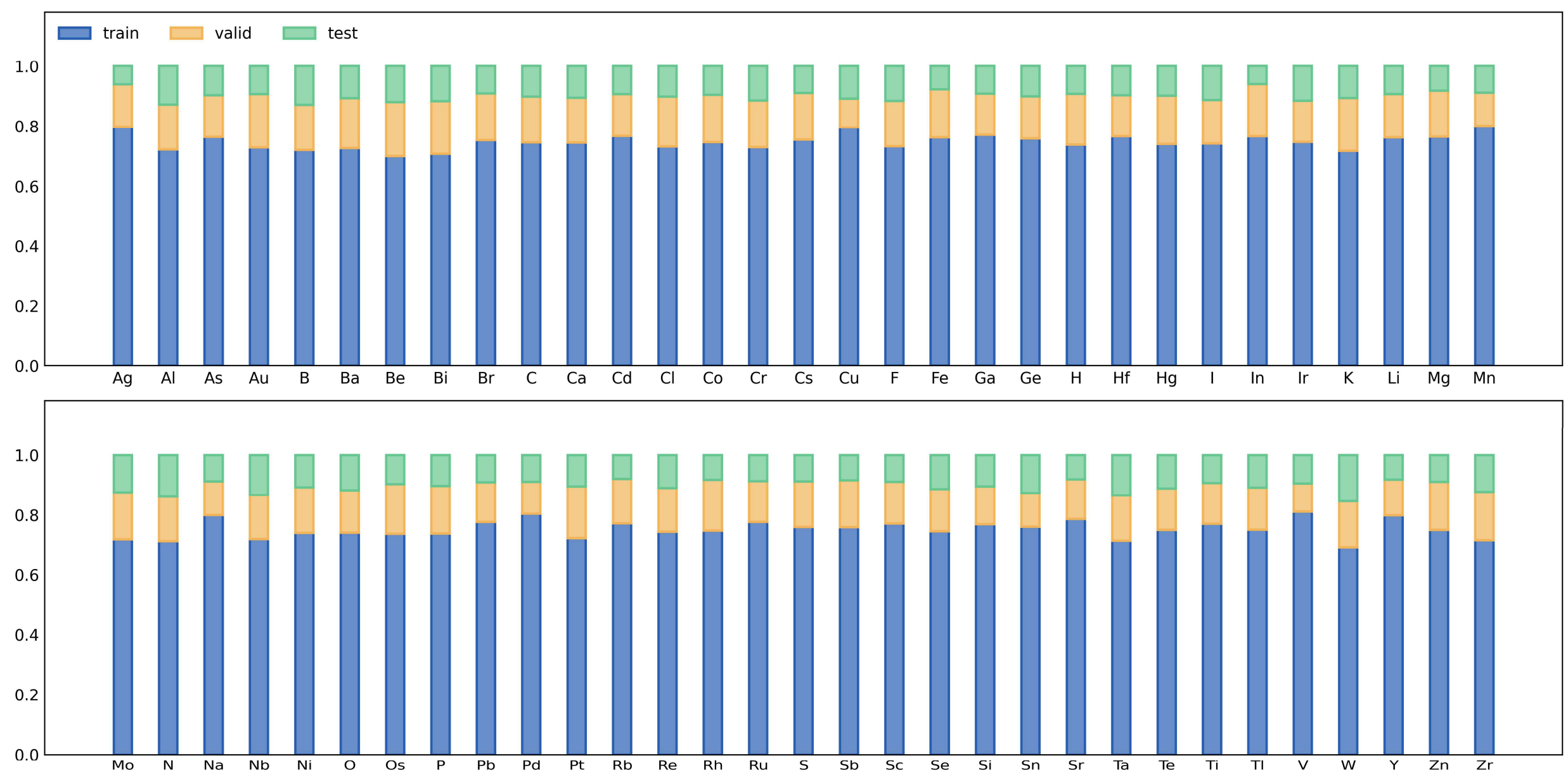

**Figure S2:** The training/validation/testing fraction split for the total number of materials that contain specific elements. Even element distribution across the training/validation/testing datasets indicates that all datasets are well-balanced.

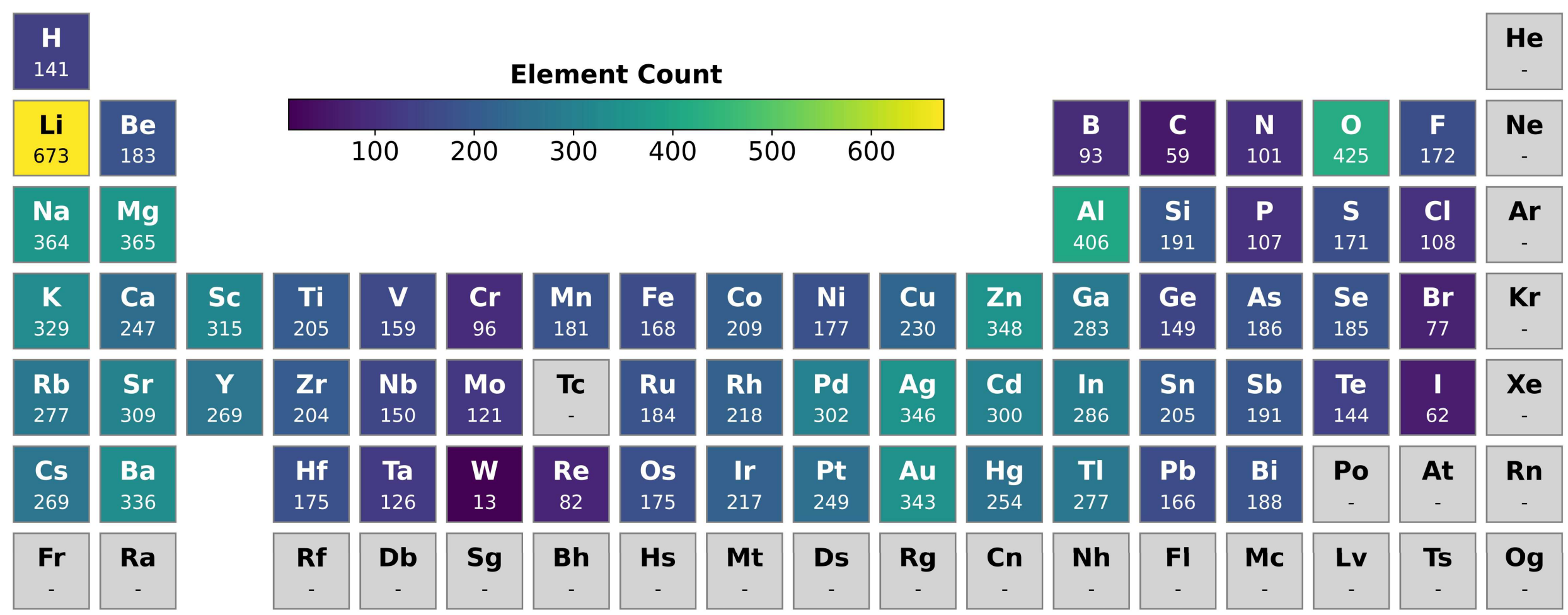

**Figure S3:** Number of materials that contain specific elements in periodic table format.

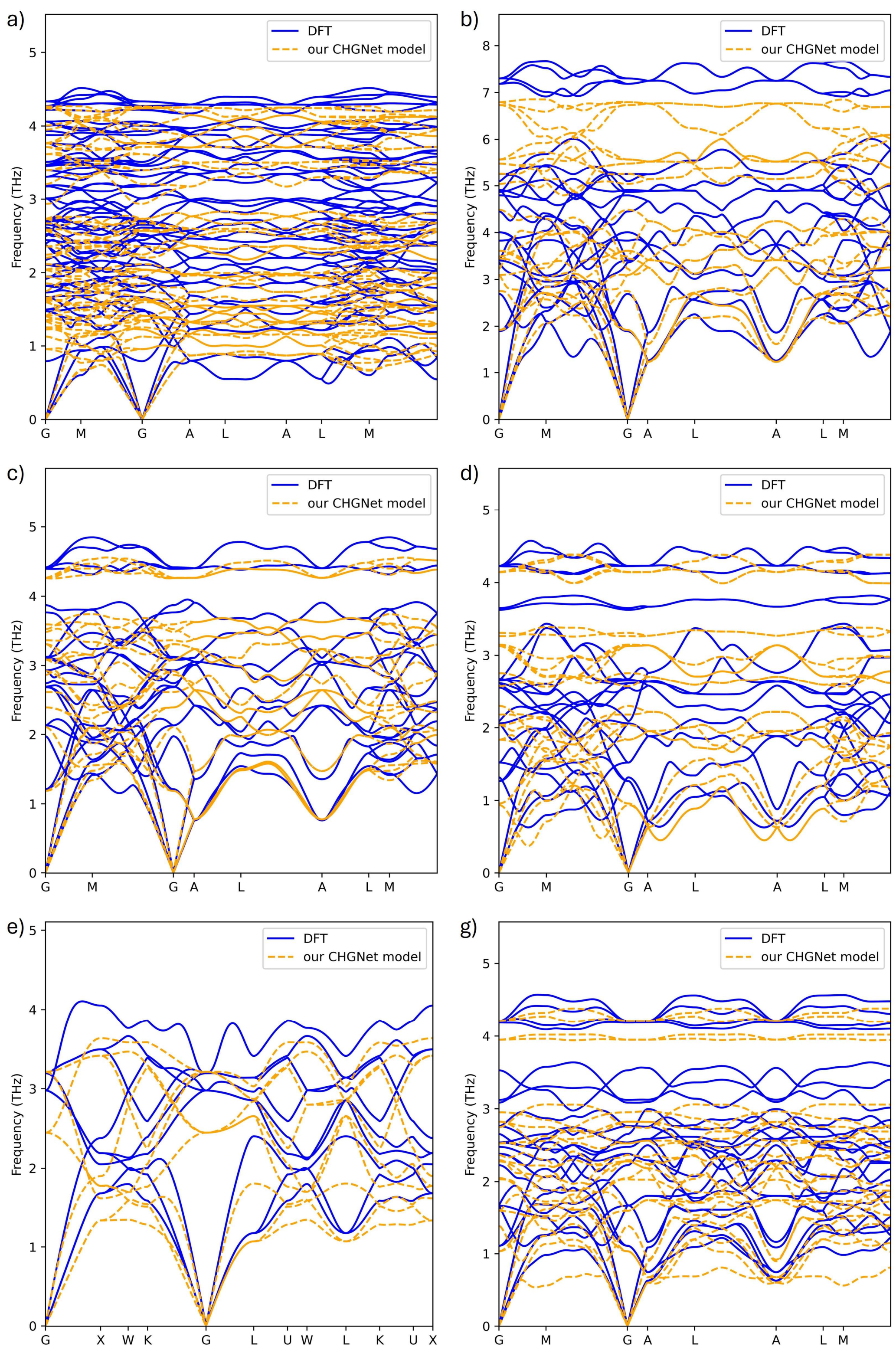

**Figure S4:** Phonon dispersion plots comparison between DFT and our CHGNet model for the following OQMD ID with corresponding chemical formula: (a) 1338889, $Sr_7Ge_3$, (b) 1366391,

$Na_2MgGe$, (c) 1369362, $K_2NaSb$, (d) 1372624, $Rb_2NaAs$, (e) 1473988, $K_2Te$, (f) 1610487, KRbNaSb.

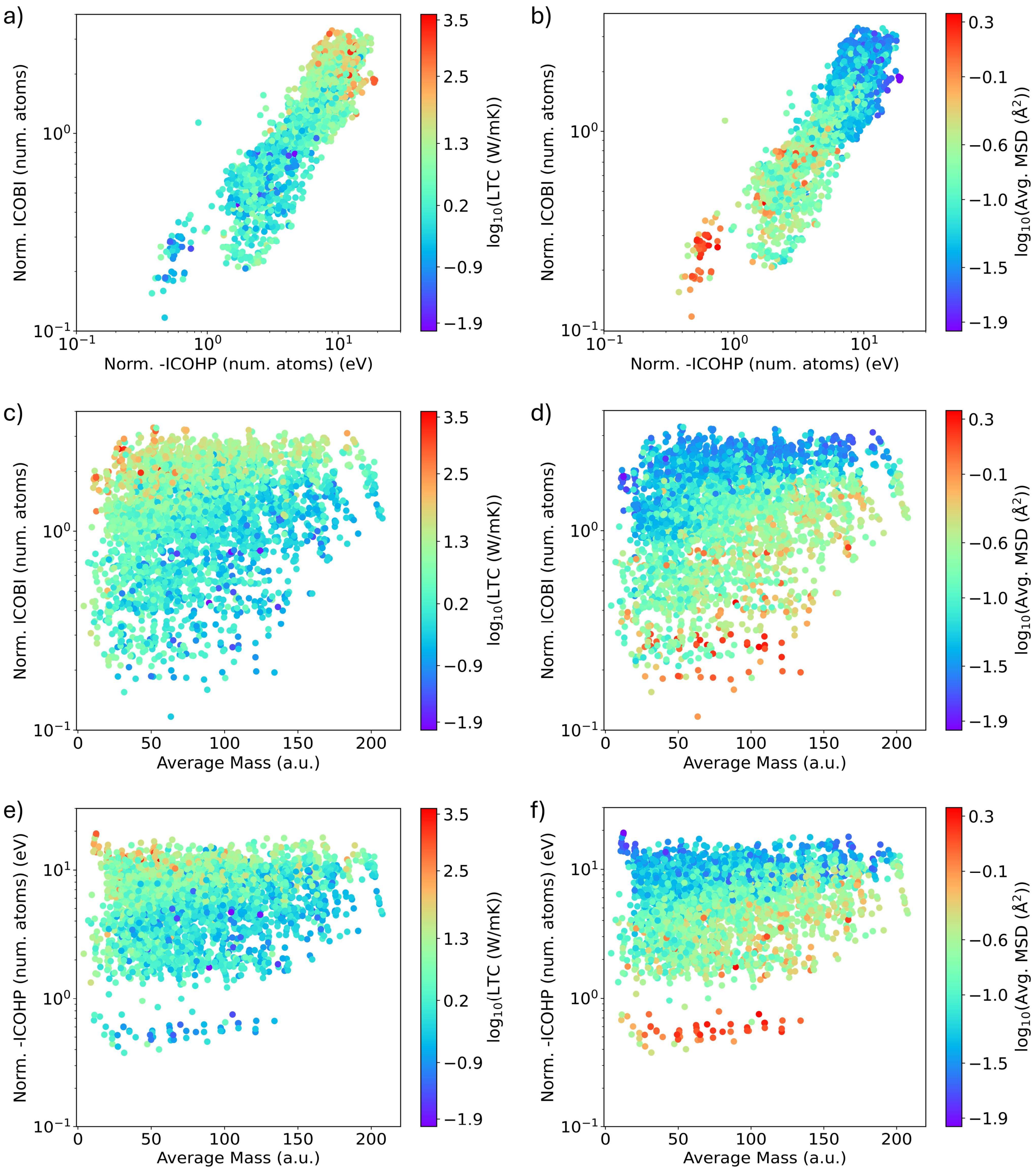

**Figure S5:** Plots for normalized -ICOHP (number of atoms) and normalized ICOBI (number of atoms) to analyze for LTC and average MSD. The plots show a) normalized ICOBI (number of atoms) and normalized -ICOHP (number of atoms) color-mapped with LTC, b) normalized ICOBI (number of atoms) and normalized -ICOHP (number of atoms) color-mapped with the average MSD. c) normalized ICOBI (number of atoms) and average mass color-mapped with LTC. d) normalized ICOBI (number of atoms) and average mass color-mapped with the average

MSD. e) normalized -ICOHP (number of atoms) and average mass color-mapped with LTC. f) normalized -ICOHP (number of atoms) and average mass color-mapped with the average MSD.

## Reference

[1] Vedernikov, M. V. The Thermoelectric Powers of Transition Metals at High Temperature. Advances in Physics 1969, 18 (74), 337–370.